\documentclass[12pt]{article}
\usepackage{amsmath,amssymb}
\usepackage{bm}
\usepackage{graphicx}

\allowdisplaybreaks[3]

\newcommand{\eqdef}{\stackrel{\rm def}{=}}

\newcommand{\n}{\nonumber \\}
\newcommand{\ignore}[1]{}

\newcommand{\maprightu}[2]
  {\smash{\mathop{\hbox to #1{\rightarrowfill}}\limits^{#2}}}
\newcommand{\maprightd}[2]
  {\smash{\mathop{\hbox to #1{\rightarrowfill}}\limits_{#2}}}

\newcommand{\mapdownr}[1]
  {\Bigg\downarrow\rlap{$\vcenter{\hbox{$\scriptstyle#1$}}$}}

\begin{document}

\baselineskip=20pt

\newfont{\elevenmib}{cmmib10 scaled\magstep1}
\newcommand{\preprint}{
  \begin{flushright}\normalsize  \sf
       DPSU-05-1\\
      {\tt hep-th/0505070}\\
       May 2005
     \end{flushright}}
\newcommand{\Title}[1]{{\baselineskip=26pt
     \begin{center} \Large \bf #1 \\ \ \\ \end{center}}}
\newcommand{\Author}{\begin{center}
     \large \bf Satoru Odake${}^a$ and Ryu Sasaki${}^b$ \end{center}}
\newcommand{\Address}{\begin{center}
       $^a$ Department of Physics, Shinshu University,\\
       Matsumoto 390-8621, Japan\\
       ${}^b$ Yukawa Institute for Theoretical Physics,\\
       Kyoto University, Kyoto 606-8502, Japan
     \end{center}}
\newcommand{\Accepted}[1]{\begin{center}
     {\large \sf #1}\\ \vspace{1mm}{\small \sf Accepted for Publication}
     \end{center}}

\preprint
\thispagestyle{empty}
\bigskip\bigskip\bigskip

\Title{Equilibrium Positions and Eigenfunctions of \\
Shape Invariant (`Discrete') Quantum Mechanics\footnote{
Based on the talk ``Equilibria of `Discrete' Integrable Systems and
Deformations of Classical Orthogonal Polynomials''
at the RIMS workshop ``Elliptic Integrable Systems''\llap, 8--11
November 2004.\\ 
Submitted to Rokko Lectures in Mathematics (Kobe University).} }
\Author

\Address
\vspace{1cm}

\begin{abstract}
Certain aspects of the integrability/solvability of the 
Calogero-Sutherland-Moser systems and the Ruijsenaars-Schneider-van Diejen
systems with rational and trigonometric potentials are reviewed.
The equilibrium positions of classical multi-particle systems
and the eigenfunctions of single-particle quantum mechanics are
described by the same orthogonal polynomials: the Hermite, Laguerre,
Jacobi, continuous Hahn, Wilson and Askey-Wilson polynomials.
The Hamiltonians of these single-particle quantum mechanical systems
have two remarkable properties, factorization and shape invariance.
\end{abstract}

\newpage
\section{Introduction}

Exactly solvable or quasi-exactly solvable multi-particle
quantum mechanical systems have many remarkable properties.
Especially, those of the Calogero-Sutherland-Moser(CSM)
systems\cite{C71,S72,CM75} and their integrable deformation called the
Ruijsenaars-Schneider-van Diejen (RSvD) systems \cite{RS86,vD94} have
been well studied.
For example, the spectral curves of the classical elliptic CSM systems
appear in the Seiberg-Witten theory of the supersymmetric gauge
theory \cite{SW94}, and the relation between the eigenstates of the
quantum Sutherland system and those of the Ruijsenaars-Schneider system
has led to the discovery of the deformed Virasoro and $W_N$ algebras 
\cite{AKOS96}.

The equilibrium positions of the Calogero-Sutherland systems are
described by the zeros of the classical orthogonal polynomials;
the Hermite, Laguerre and Jacobi (Chebyshev, Legendre, Gegenbauer)
polynomials \cite{C77,Sti,Sze,OS1}.
Motivated by the simple reasoning illustrated in the following diagram
(a similar idea has led to the deformed Virasoro and $W_N$ algebras),
\medskip
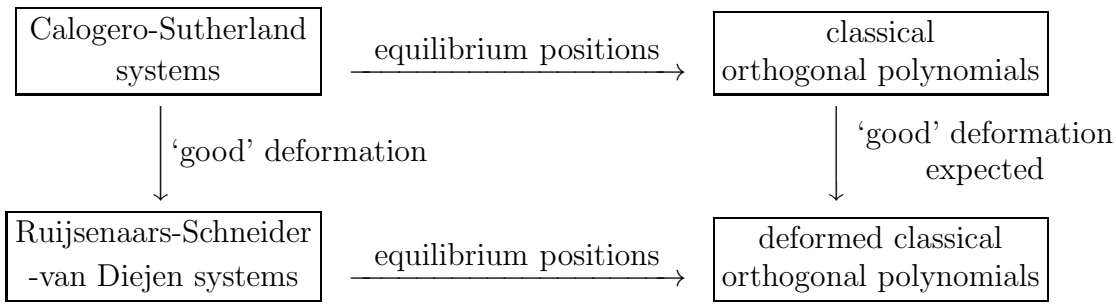
\begin{eqnarray*}
   \framebox{\shortstack{$\,$Calogero-Sutherland$\,$ \\ \\ systems}}
   &\maprightu{45mm}{\mbox{equilibrium positions}}&
   \framebox{\shortstack{classical \\ \\
       orthogonal polynomials}}\\
   \mapdownr{\mbox{`good' deformation}}\hspace{20mm}&&
   \hspace{15mm}\mapdownr{\makebox{
      \shortstack{`good' deformation \\ expected}}}\\
   \hspace*{-5mm}
   \framebox{\shortstack{Ruijsenaars-Schneider \\ \\ -van Diejen systems}}
   &\maprightu{45mm}{\mbox{equilibrium positions}}&
   \framebox{\shortstack{deformed classical \\ \\
            orthogonal polynomials}}
\end{eqnarray*}
we studied the equilibrium positions of the RSvD systems with rational
and trigonometric potentials by using numerical analysis, functional
equation and three-term recurrence.
This program worked well and we obtained the deformed polynomials
\cite{OS2,OS5} (see also \cite{RagS04,vD04,ILR04}),
which fitted in the Askey-scheme of the hypergeometric orthogonal
polynomials \cite{AW85,KS96,AAR99}:
deformation of the Hermite polynomial $\Rightarrow$ special cases of the
Meixner-Pollaczek polynomial and the continuous Hahn polynomial;
deformation of the Laguerre polynomial $\Rightarrow$
the continuous dual Hahn polynomial and the Wilson polynomial;
deformation of the Jacobi polynomial $\Rightarrow$
the Askey-Wilson polynomial.

The Hermite, Laguerre and Jacobi polynomials, which describe the
equilibrium positions of the classical multi-particle CS systems, also
describe the eigenfunctions of the corresponding single-particle quantum
CS systems. This interesting property is inherited by the deformed ones.
The continuous Hahn, Wilson and Askey-Wilson polynomials, which
describe the equilibrium positions of the classical multi-particle RSvD
systems, also describe the eigenfunctions of the corresponding 
single-particle quantum RSvD systems.
The Hamiltonians for single-particle quantum CS and RSvD systems have
two remarkable properties, {\em factorization} and {\em shape invariance}
\cite{IH51,Crum55,Genden,SVZ93,susyqm}.
Shape invariance is an important ingredient of many exactly solvable
quantum mechanics.
In our case the shape invariance determines the eigenfunctions and spectrum
from the data of the ground state wavefunction and the energy of the
first excited state \cite{OS4,OS5}.

The aim of this note is to give a comprehensive review of the above
facts; (a) equilibrium positions and single-particle eigenfunctions of
the CS and RSvD systems with rational and trigonometric potentials are 
described by the same orthogonal polynomials, 
(b) Hamiltonians for quantum single-particle CS and RSvD systems with
rational and trigonometric potentials have two properties, factorization
and shape invariance.

This note is organized as follows.
In section 2 we recapitulate the essence of the models, CS and RSvD
systems with rational and trigonometric potentials. The relationship
between CS and RSvD systems is discussed, and the equations for the 
equilibrium positions are given.
In sections 3--6 we demonstrate that the same polynomials appear in the
equilibrium positions and single-particle eigenfunctions.
We emphasize the shape invariance of the single-particle Hamiltonian
along the idea of Crum\cite{Crum55}.
In section 3 the CS and RSvD systems with rational $A$-type potentials
are discussed and the relevant polynomials are the Hermite and the
continuous Hahn polynomials.
In section 4 rational $BC$-type potentials are discussed and the
Laguerre and Wilson polynomials play the role.
Section 5 is for the trigonometric $A$-type potentials.
In section 6 the trigonometric $BC$-type potentials are discussed and
the Jacobi and Askey-Wilson polynomials appear.
Section 7 is for summary and comments.

\section{Models}

We recapitulate the basics of  the models and present the
equations for their equilibrium positions.
A multi-particle quantum (or classical) mechanics governed
by a Hamiltonian $H(p,q)$ (or classical one $H^{\rm class}(p,q)$)
is considered.
The dynamical variables are real-valued coordinates
$q={}^t(q_1,\cdots,q_n)$ and their canonically conjugate momenta
$p={}^t(p_1,\cdots,p_n)$.
For quantum case we have $p_j=-i\hbar\frac{\partial}{\partial q_j}$.
We keep dimensionful parameters, e.g. mass, angular frequency, the
Planck constant, etc. The coordinate $q_j$ has dimension of length.

\subsection{Calogero-Sutherland Systems}

The Hamiltonian of the Calogero-Sutherland (CS) systems is
\begin{equation}
  H_{\rm CS}(p,q)=\sum_{j=1}^n\frac{1}{2m}p_j^2+V_{\rm CS}(q)\,,
  \label{H_CS}
\end{equation}
where the potential $V_{\rm CS}(q)$ can be written in terms of the
prepotential $W(q)$,
\begin{equation}
  V_{\rm CS}(q)=\sum_{j=1}^n\frac{1}{2m}\biggl(
  \Bigl(\frac{\partial W(q)}{\partial q_j}\Bigr)^2
  +\hbar\,\frac{\partial^2 W(q)}{\partial q_j^2}\biggr)\,.
  \label{V=W^2+dW}
\end{equation}
The explicit forms of the potential $V_{\rm CS}(q)$ and the prepotential
$W(q)$ are as follows:\\ 
(\romannumeral1) rational $A_{n-1}$ :
\begin{align}
  V_{\rm CS}(q)&=\sum_{j=1}^n\frac12m\omega^2q_j^2
  +\frac{\hbar^2}{2m}\sum_{\genfrac{}{}{0pt}{2}{j,k=1}{j\neq k}}^n
  \frac{g(g-1)}{(q_j-q_k)^2}
  -\tfrac12\hbar\omega n\bigl(1+g(n-1)\bigr)\,,
  \label{ratA_CS_V}\\
  W(q)&=-\sum_{j=1}^n\frac12m\omega q_j^2+\sum_{1\leq j<k\leq n}
  g\hbar\log\sqrt{\tfrac{m\omega}{\hbar}}\,\bigl|q_j-q_k\bigr|\,,
  \label{ratA_CS_W}
\end{align}
(\romannumeral2) rational\footnote{
Since the independent coupling constants are $g_M$ and $g_S+g_L$,
this $BC_n$ model is equivalent to $B_n$ or $C_n$ model.
} $BC_n$ :
\begin{align}
  V_{\rm CS}(q)&=\sum_{j=1}^n\biggl(\frac12m\omega^2q_j^2
  +\frac{\hbar^2}{2m}\frac{(g_S+g_L)(g_S+g_L-1)}{q_j^2}\biggr)\n
  &\quad
  +\frac{\hbar^2}{2m}\sum_{\genfrac{}{}{0pt}{2}{j,k=1}{j\neq k}}^n
  \biggl(\frac{g_M(g_M-1)}{(q_j-q_k)^2}+\frac{g_M(g_M-1)}{(q_j+q_k)^2}
  \biggr)
  -\hbar\omega n\bigl(g_S+g_L+\tfrac12+g_M(n-1)\bigr)\,,
  \label{ratBC_CS_V}\\
  W(q)&=-\sum_{j=1}^n\frac12m\omega q_j^2+\sum_{1\leq j<k\leq n}
  g_M\hbar\Bigl(\log\sqrt{\tfrac{m\omega}{\hbar}}\,\bigl|q_j-q_k\bigr|
  +\log\sqrt{\tfrac{m\omega}{\hbar}}\,\bigl|q_j+q_k\bigr|\Bigr)\n
  &\quad+\sum_{j=1}^n\Bigl(
  g_S\hbar\log\sqrt{\tfrac{m\omega}{\hbar}}\,\bigl|q_j\bigr|
  +g_L\hbar\log\sqrt{\tfrac{m\omega}{\hbar}}\,\bigl|2q_j\bigr|\Bigr)\,,
  \label{ratBC_CS_W}
\end{align}
(\romannumeral3) trigonometric $A_{n-1}$ :
\begin{align}
  V_{\rm CS}(q)&=
  \frac{\hbar^2\pi^2}{2mL^2}\sum_{\genfrac{}{}{0pt}{2}{j,k=1}{j\neq k}}^n
  \frac{g(g-1)}{\sin^2\frac{\pi}{L}(q_j-q_k)}
  -\frac{\hbar^2\pi^2}{2mL^2}g^2\frac13n(n^2-1)\,,
  \label{trigA_CS_V}\\
  W(q)&=\sum_{1\leq j<k\leq n}g\hbar
  \log\bigl|\sin\tfrac{\pi}{L}(q_j-q_k)\bigr|\,,
  \label{trigA_CS_W}
\end{align}
(\romannumeral4) trigonometric $BC_n$ :
\begin{align}
  V_{\rm CS}(q)&=\frac{\hbar^2\pi^2}{2mL^2}\sum_{j=1}^n\biggl(
  \frac{(g_S+g_L)(g_S+g_L-1)}{\sin^2\frac{\pi}{L}q_j}
  +\frac{g_L(g_L-1)}{\cos^2\frac{\pi}{L}q_j}\biggl)\n
  &\quad
  +\frac{\hbar^2\pi^2}{2mL^2}\sum_{\genfrac{}{}{0pt}{2}{j,k=1}{j\neq k}}^n
  \biggl(\frac{g_M(g_M-1)}{\sin^2\frac{\pi}{L}(q_j-q_k)}
  +\frac{g_M(g_M-1)}{\sin^2\frac{\pi}{L}(q_j+q_k)}\biggr)\n
  &\quad
  -\frac{\hbar^2\pi^2}{2mL^2}n\Bigl(\bigl(g_S+2g_L+g_M(n-1)\bigr)^2
  +g_M^2\tfrac13(n^2-1)\Bigr)\,,
  \label{trigBC_CS_V}\\
  W(q)&=\sum_{1\leq j<k\leq n}g_M\hbar\Bigl(
  \log\bigl|\sin\tfrac{\pi}{L}(q_j-q_k)\bigr|
  +\log\bigl|\sin\tfrac{\pi}{L}(q_j+q_k)\bigr|\Bigr)\n
  &\quad
  +\sum_{j=1}^n\Bigl(g_S\hbar\log\bigl|\sin\tfrac{\pi}{L}q_j\bigr|
  +g_L\hbar\log\bigl|\sin\tfrac{\pi}{L}2q_j\bigr|\Bigr)\,.
  \label{trigBC_CS_W}
\end{align}
The constant terms in $V_{\rm CS}(q)$ are the consequences of the
expression (\ref{V=W^2+dW}) in terms of the prepotential.
A constant shift of $W(q)$ does not affect \eqref{V=W^2+dW}.
In those formulas $g,g_S,g_M$ and $g_L$ are dimensionless couplings
constants and we assume they are positive.
The other notation is conventional;
$m$ is the mass of particles, $\omega$ is the angular frequency,
$\hbar$ is the Planck constant (divided by $2\pi$) and  $L$ is the
circumference. All these parameters are positive.

\subsection{Ruijsenaars-Schneider-van Diejen Systems}

The Ruijsenaars-Schneider-van Diejen (RSvD) systems are deformation of
the Calogero-Sutherland-Moser systems.
The Hamiltonian of RSvD systems is
\begin{equation}
  H(p,q)=\frac12mc^2\sum_{j=1}^n\Bigl(
  \sqrt{V_j(q)}\,e^{\frac{1}{mc}p_j}\sqrt{{V_j(q)}^*}
  +\sqrt{{V_j(q)}^*}\,e^{-\frac{1}{mc}p_j}\sqrt{V_j(q)}
  -V_j(q)-{V_j(q)}^*\Bigr)\,,
  \label{H_RS}
\end{equation}
where $V_j(q)$ are
\begin{equation}
  V_j(q)=w(q_j)\prod_{\genfrac{}{}{0pt}{2}{k=1}{k\neq j}}^n
  v(q_j-q_k)\times
  \begin{cases}
  1&\text{for $A_{n-1}$}\\v(q_j+q_k)&\text{for $BC_n$}\,.
  \end{cases}
\end{equation}
Since operators $e^{\pm\frac{1}{mc}p_j}=
e^{\mp i\frac{\hbar}{mc}\frac{\partial}{\partial q_j}}$ cause finite
shifts of the wavefunction in the imaginary direction
($e^{\pm\frac{1}{mc}p_j}f(q)=
f(q_1,\cdots,q_j\mp i\frac{\hbar}{mc},\cdots,q_n)$),
we call these systems `discrete' dynamical systems.\footnote{
Sometimes they are misleadingly called `relativistic' version of the CSM.
See \cite{BS97} for comments on this point.
}
The basic potential functions $v(x)$ and $w(x)$ are given by as follows:\\
(\romannumeral1) rational $A_{n-1}$ :
\begin{align}
  v(x)&=1-i\frac{\hbar}{mc}\frac{g}{x}\,,
  \label{ratA_RS_v}\\
  w(x)&=\Bigl(1+i\,\frac{\omega_1}{c}\,x\Bigr)
  \Bigl(1+i\,\frac{\omega_2}{c}\,x\Bigr)\,,
  \label{ratA_RS_w}
\end{align}
(\romannumeral2) rational $BC_n$ :
\begin{align}
  v(x)&=1-i\frac{\hbar}{mc}\frac{g_0}{x}\,,
  \label{ratBC_RS_v}\\
  w(x)&=\Bigl(1+i\,\frac{\omega_1}{c}\,x\Bigr)
  \Bigl(1+i\,\frac{\omega_2}{c}\,x\Bigr)
  \Bigl(1-i\frac{\hbar}{mc}\frac{g_1}{x}\Bigr)
  \Bigl(1-i\frac{\hbar}{mc}\frac{g_2}{x-i\frac{\hbar}{2mc}}\Bigr)\,,
  \label{ratBC_RS_w}
\end{align}
(\romannumeral3) trigonometric $A_{n-1}$ :
\begin{align}
  v(x)&=\frac{\sin\frac{\pi}{L}(x-i\frac{\hbar}{mc}g)}{\sin\frac{\pi}{L}x}
  \,,
  \label{trigA_RS_v}\\
  w(x)&=1\,,
  \label{trigA_RS_w}
\end{align}
(\romannumeral4) trigonometric $BC_n$ :
\begin{align}
  v(x)&=\frac{\sin\frac{\pi}{L}(x-i\frac{\hbar}{mc}g_0)}{\sin\frac{\pi}{L}x}
  \,,
  \label{trigBC_RS_v}\\
  w(x)&=
  \frac{\sin\frac{\pi}{L}(x-i\frac{\hbar}{mc}g_1)}{\sin\frac{\pi}{L}x}\,
  \frac{\sin\frac{\pi}{L}(x-i\frac{\hbar}{2mc}-i\frac{\hbar}{mc}g_2)}
       {\sin\frac{\pi}{L}(x-i\frac{\hbar}{2mc})}\n
  &\quad\times
  \frac{\cos\frac{\pi}{L}(x-i\frac{\hbar}{mc}g'_1)}{\cos\frac{\pi}{L}x}\,
  \frac{\cos\frac{\pi}{L}(x-i\frac{\hbar}{2mc}-i\frac{\hbar}{mc}g'_2)}
       {\cos\frac{\pi}{L}(x-i\frac{\hbar}{2mc})}\,.
  \label{trigBC_RS_w}
\end{align}
Here $g,g_0,g_1,g_2,g'_1$ and $g'_2$ are dimensionless couplings
constants and $c$ is the (fictitious) speed of light. 
We assume they are all positive.

Let us consider $c\rightarrow\infty$ limit, in which RSvD systems reduce
to CS systems.
Since $v(x)$ and $w(x)$ contains $c$ and $i$ as a combination
$\frac{i}{c}$, we can expand them as follows:
\begin{align}
  v(x)&=1+\tfrac{i}{c}\,v_1(x)+(\tfrac{i}{c})^2v_2(x)+O(\tfrac{1}{c^3})\,,\\
  w(x)&=1+\tfrac{i}{c}\,w_1(x)+(\tfrac{i}{c})^2w_2(x)+O(\tfrac{1}{c^3})\,.
\end{align}
Here $v_1(x)$ and $w_1(x)$ are odd real functions and $v_2(x)$ and $w_2(x)$
are even real functions because of $v(x)^*=v(-x)$ and $w(x)^*=w(-x)$.
Then the Hamiltonian \eqref{H_RS} has the expansion,
\begin{align}
  H(p,q)&=\sum_{j=1}^n\frac{1}{2m}p_j^2
  +\sum_{j=1}^n\Bigl(\frac{m}{2}w_1(q_j)^2-\frac{\hbar}{2}w'_1(q_j)\Bigr)\n
  &\quad+\sum_{\genfrac{}{}{0pt}{2}{j,k=1}{j\neq k}}^n
  \Bigl(\frac{m}{2}v_1(q_j-q_k)^2-\frac{\hbar}{2}v'_1(q_j-q_k)
  +\frac{m}{2}v_1(q_j+q_k)^2-\frac{\hbar}{2}v'_1(q_j+q_k)\n
  &\qquad\qquad\quad+mw_1(q_j)\bigl(v_1(q_j-q_k)+v_1(q_j+q_k)\bigr)\Bigr)\n
  &\quad+\sum_{\genfrac{}{}{0pt}{2}{j,k,l=1}{j\neq k\neq l\neq j}}^n
  \frac{m}{2}\bigl(v_1(q_j-q_k)+v_1(q_j+q_k)\bigr)
  \bigl(v_1(q_j-q_l)+v_1(q_j+q_l)\bigr)
  +O\Bigl(\frac{1}{c}\Bigr)\,,
\end{align}
where the prime stands for the derivative. (For $A_{n-1}$ type systems,
the terms containing $v_1(q_j+q_k)$ and $v'_1(q_j+q_k)$ should be omitted.)
By explicit calculation, we obtain
\begin{equation}
  \lim_{c\rightarrow\infty}H(p,q)=H_{\rm CS}(p,q),
  \label{RSCS}
\end{equation}
where the correspondence of parameters are
\begin{align}
  \text{(\romannumeral1)}\quad&
  \omega_1+\omega_2=\omega,\quad g=g,
  \label{ratA_RSCS_para}\\
  \text{(\romannumeral2)}\quad&
  \omega_1+\omega_2=\omega,\quad g_0=g_M,\quad g_1+g_2=g_S+g_L,
  \label{ratBC_RSCS_para}\\
  \text{(\romannumeral3)}\quad&
  g=g,
  \label{trigA_RSCS_para}\\
  \text{(\romannumeral4)}\quad&
  g_0=g_M,\quad g_1+g_2=g_S+g_L,\quad g'_1+g'_2=g_L\,.
  \label{trigBC_RSCS_para}
\end{align}

\subsection{Equilibrium Positions}

The classical Hamiltonian $H^{\rm class}(p,q)$ is obtained from the quantum
one $H(p,q)$ by the following procedure; (a) regard $p_j$ is a $c$-number,
(b) after expressing dimensionless coupling constants $g,g_1,g_2,\cdots$ by
dimensionful coupling constants $\bar{g}=g\hbar, {\bar{g}}_1=g_1\hbar,
{\bar{g}}_2=g_2\hbar,\cdots$, assume
$\bar{g},{\bar{g}}_1,{\bar{g}}_2,\cdots$ are independent of $\hbar$,
(c) take $\hbar\rightarrow0$ limit.
In the same way, $V^{\rm class}(q)$, $W^{\rm class}(q)$,
$V_j^{\rm class}(q)$, $v^{\rm class}(x)$ and $w^{\rm class}(x)$ are also
obtained.

The canonical equations of motion of the classical systems are
\begin{equation}
  \frac{dq_j}{dt}=\frac{\partial H^{\rm class}(p,q)}{\partial p_j}\,,\quad
  \frac{dp_j}{dt}=-\frac{\partial H^{\rm class}(p,q)}{\partial q_j}\,.
\end{equation}
The equilibrium positions are the stationary solution
\begin{equation}
  p=0\,,\quad q=\bar{q}\,,
  \label{stasol}
\end{equation}
in which $\bar{q}$ satisfies
\begin{equation}
  \frac{\partial H^{\rm class}(0,q)}{\partial q_j}\Biggm|_{q=\bar{q}}=0
  \qquad(j=1,\ldots,n).
  \label{staeq}
\end{equation}

For CS system, \eqref{staeq} becomes
$\frac{\partial V^{\rm class}_{\rm CS}(q)}{\partial q_j}\Bigm|_{q=\bar{q}}=0$
and it is equivalent to the condition \cite{CorS02}
\begin{equation}
  \frac{\partial W^{\rm class}(q)}{\partial q_j}\Biggm|_{q=\bar{q}}=0
  \qquad(j=1,\ldots,n).
  \label{CS_equiveqW}
\end{equation}
For RSvD system, \eqref{staeq} is equivalent to the condition \cite{RagS04}
\begin{equation}
  V_j^{\rm class}(\bar{q})=V_j^{\rm class}(\bar{q})^*>0\qquad(j=1,\ldots,n).
  \label{RS_equiveq}
\end{equation}
This equation {\em without inequality\/} is rewritten in a Bethe
ansatz like equation
\begin{equation}
  \prod_{\genfrac{}{}{0pt}{2}{k=1}{k\neq j}}^n
  \frac{v^{\rm class}(\bar{q}_j-\bar{q}_k)\,v^{\rm class}(\bar{q}_j+\bar{q}_k)}
  {v^{\rm class}(\bar{q}_j-\bar{q}_k)^*\,v^{\rm class}(\bar{q}_j+\bar{q}_k)^*}
  =\frac{w^{\rm class}(\bar{q}_j)^*}{w^{\rm class}(\bar{q}_j)}
  \qquad(j=1,\ldots,n).
  \label{equiveqBA}
\end{equation}
(For $A_{n-1}$ type systems, $v^{\rm class}(\bar{q}_j+\bar{q}_k)$ and
$v^{\rm class}(\bar{q}_j+\bar{q}_k)^*$ should be omitted.)

\section{Rational $A$ Types}

In this section we consider CS and RSvD systems with rational $A$ type
potentials.
Relevant polynomials are the Hermite polynomial and the continuous Hahn
polynomial.

\subsection{Calogero Systems}

The Hamiltonian is \eqref{H_CS} with the potential
\eqref{ratA_CS_V}--\eqref{ratA_CS_W}.

\subsubsection{Equilibrium positions of $n$-particle classical systems}

For the $n$-particle prepotential \eqref{ratA_CS_W}, the equation for the
equilibrium positions \eqref{CS_equiveqW} was studied by Stieltjes in a
slightly different context more than a century ago \cite{Sti}.
Let us consider a polynomial whose zeros give the equilibrium positions,
$f(y)=\prod_{j=1}^n(y-\sqrt{\frac{m\omega}{\bar{g}}}\,\bar{q}_j)$.
Then \eqref{CS_equiveqW} can be converted to a differential equation
for $f(y)$, which is the determining  equation for the Hermite
polynomial. Therefore we obtain the result \cite{C77},
\begin{equation}
  \prod_{j=1}^n\Bigl(y-\sqrt{\frac{m\omega}{\bar{g}}}\,\bar{q}_j\Bigr)
  =H_n^{\rm monic}(y)\,,
\end{equation}
where $H_n(y)=2^nH_n^{\rm monic}(y)$ is the Hermite polynomial \cite{KS96}.

\subsubsection{Eigenfunctions of single-particle quantum mechanics}
\label{1particle_ratA_CS}

Let us consider single-particle case ($n=1$) and write $x=q_1$.
The Hamiltonian \eqref{H_CS} describes the harmonic oscillator with
the constant energy shift
\begin{equation}
  H=-\frac{\hbar^2}{2m}\frac{d^2}{dx^2}+\frac12m\omega^2x^2
  -\frac12\hbar\omega\,.
\end{equation}
The eigenfunctions of this Hamiltonian are well-known, but we describe
it in detail in order to illustrate the idea of Crum\cite{Crum55}, 
{\em construction of isospectral Hamiltonians} (see Figure 1).
By introducing a dimensionless variable $y$,
\begin{equation}
  y=\sqrt{\frac{m\omega}{\hbar}}x,
  \label{rat_CS_y}
\end{equation}
$H$ can be written as
\begin{equation}
  H=\hbar\omega\mathcal{H}\,,\quad
  \mathcal{H}=-\frac12\frac{d^2}{dy^2}+\frac12y^2-\frac12\,.
  \label{ratA_CS_cH}
\end{equation}
Instead of $H\phi_n=E_n\phi_n$, let us consider a rescaled one
$\mathcal{H}\phi_n(y)=\mathcal{E}_n\phi_n(y)$ ($n=0,1,2,\ldots$),
where energies are related as $E_n=\hbar\omega\mathcal{E}_n$.

Let us develop the factorization method in its fullest generality.
Let us assume that a single-particle Hamiltonian $\mathcal{H}$ depends
on a set of parameters to be represented collectively as $\bm{\lambda}$.
The present Hamiltonian $\mathcal{H}$ \eqref{ratA_CS_cH} contains no
parameter, though.
The Hamiltonian $\mathcal{H}$, defined in terms of the prepotential, is
factorizable:
\begin{align}
  \mathcal{H}&=\mathcal{H}(y\,;\bm{\lambda})
  =\mathcal{A}(y\,;\bm{\lambda})^{\dagger}\mathcal{A}(y\,;\bm{\lambda})
  =\frac12\Bigl(-\frac{d^2}{dy^2}
  +\Bigl(\frac{d\mathcal{W}(y\,;\bm{\lambda})}{dy}\Bigr)^2
  +\frac{d^2\mathcal{W}(y\,;\bm{\lambda})}{dy^2}\Bigr)\,,
  \label{rat_CS_calH=AA}\\
  \mathcal{A}&=\mathcal{A}(y\,;\bm{\lambda})\eqdef\frac{1}{\sqrt{2}}\Bigl(
  -i\frac{d}{dy}+i\,\frac{d\,\mathcal{W}(y\,;\bm{\lambda})}{dy}\Bigr)\,,\\
  \mathcal{A}^{\dagger}&=
  \mathcal{A}(y\,;\bm{\lambda})^{\dagger}\eqdef\frac{1}{\sqrt{2}}\Bigl(
  -i\frac{d}{dy}-i\,\frac{d\,\mathcal{W}(y\,;\bm{\lambda})}{dy}\Bigr)\,,
  \label{rat_CS_Ad}
\end{align}
where $\mathcal{W}(y\,;\bm{\lambda})$ is a prepotential.
The ground state of $\mathcal{H}$ is  annihilated by
$\mathcal{A}$ (see Remark in \S\ref{1particle_ratBC_CS}) and it is
expressed by $\mathcal{W}$,
\begin{equation}
  \phi_0(y\,;\bm{\lambda})\propto e^{\mathcal{W}(y\,;\bm{\lambda})},\quad
  \mathcal{E}_0(\bm{\lambda})=0.
  \label{rat_CS_phi0}
\end{equation}
In the present case we have
\begin{equation}
  \mathcal{W}(y)=-\frac12y^2,\quad\phi_0(y)\propto e^{-\frac12y^2},
\end{equation}
which is obviously square-integrable.

This Hamiltonian has a good property, {\em shape invariance}.
Its key identity is
\begin{equation}
  \mathcal{A}(y\,;\bm{\lambda})\mathcal{A}(y\,;\bm{\lambda})^{\dagger}
  =\mathcal{A}(y\,;\bm{\lambda}+\bm{\delta})^{\dagger}
  \mathcal{A}(y\,;\bm{\lambda}+\bm{\delta})
  +\mathcal{E}_1(\bm{\lambda})\,,
  \label{rat_CS_shapeinv}
\end{equation}
where $\bm{\delta}$ stands for a set of constants.
In the present case we have
\begin{equation}
  \mathcal{E}_1=1\,,
  \label{ratA_CS_E1}
\end{equation}
and there is no $\bm{\delta}$ because of no $\bm{\lambda}$.
Starting from $\mathcal{A}_0=\mathcal{A}$, $\mathcal{H}_0=\mathcal{H}$
and $\phi_{0,n}=\phi_n$, let us define $\mathcal{A}_s$, $\mathcal{H}_s$
and $\phi_{s,n}$ ($n\geq s\geq 0$) recursively:
\begin{align}
  \mathcal{A}_{s+1}(y\,;\bm{\lambda})&\eqdef
  \mathcal{A}_s(y\,;\bm{\lambda}+\bm{\delta})\,,
  \label{rat_CS_A_rec}\\
  \mathcal{H}_{s+1}(y\,;\bm{\lambda})&\eqdef
  \mathcal{A}_s(y\,;\bm{\lambda})\mathcal{A}_s(y\,;\bm{\lambda})^{\dagger}
  +\mathcal{E}_s(\bm{\lambda})\,,\\
  \phi_{s+1,n}(y\,;\bm{\lambda})&\eqdef
  \mathcal{A}_s(y\,;\bm{\lambda})\phi_{s,n}(y\,;\bm{\lambda})\,.
  \label{rat_CS_phi_rec}
\end{align}
As a consequence of the shape invariance \eqref{rat_CS_shapeinv},
we obtain for $n\geq s\geq 0$,
\begin{align}
  &\mathcal{A}_s(y\,;\bm{\lambda})
  =\mathcal{A}(y\,;\bm{\lambda}+s\bm{\delta})\,,
  \label{rat_CS_As}\\
  &\mathcal{H}_s(y\,;\bm{\lambda})
  =\mathcal{A}_s(y\,;\bm{\lambda})^{\dagger}\mathcal{A}_s(y\,;\bm{\lambda})
  +\mathcal{E}_s(\bm{\lambda})
  =\mathcal{H}(y\,;\bm{\lambda}+s\bm{\delta})+\mathcal{E}_s(\bm{\lambda})\,,
  \label{rat_CS_H_rec}\\
  &\mathcal{E}_{s+1}(\bm{\lambda})
  =\mathcal{E}_s(\bm{\lambda})+\mathcal{E}_1(\bm{\lambda}+s\bm{\delta})\,,
  \label{rat_CS_E_rec}\\
  &\mathcal{H}_s(y\,;\bm{\lambda})\phi_{s,n}(y\,;\bm{\lambda})
  =\mathcal{E}_n(\bm{\lambda})\phi_{s,n}(y\,;\bm{\lambda})\,,\\
  &\mathcal{A}_s(y\,;\bm{\lambda})\phi_{s,s}(y\,;\bm{\lambda})=0\,,\\
  &\mathcal{A}_s(y\,;\bm{\lambda})^{\dagger}\phi_{s+1,n}(y\,;\bm{\lambda})
  =\bigl(\mathcal{E}_n(\bm{\lambda})-\mathcal{E}_s(\bm{\lambda})\bigr)
  \phi_{s,n}(y\,;\bm{\lambda})\,.
  \label{rat_CS_Adphi}
\end{align}
{}From \eqref{rat_CS_phi_rec} and \eqref{rat_CS_Adphi} we obtain
formulas relating the wavefunctions along the horizontal line (the 
{\em isospectral line}) of Fig.1,
\begin{align}
  \phi_{s,n}(y\,;\bm{\lambda})&=\mathcal{A}_{s-1}(y\,;\bm{\lambda})\cdots
  \mathcal{A}_1(y\,;\bm{\lambda})\mathcal{A}_0(y\,;\bm{\lambda})
  \phi_n(y\,;\bm{\lambda})\,,
  \label{rat_CS_phi=Aphi}\\
  \phi_n(y\,;\bm{\lambda})&=
  \frac{\mathcal{A}_0(y\,;\bm{\lambda})^{\dagger}}
       {\mathcal{E}_n(\bm{\lambda})-\mathcal{E}_0(\bm{\lambda})}\,
  \frac{\mathcal{A}_1(y\,;\bm{\lambda})^{\dagger}}
       {\mathcal{E}_n(\bm{\lambda})-\mathcal{E}_1(\bm{\lambda})}\cdots
  \frac{\mathcal{A}_{n-1}(y\,;\bm{\lambda})^{\dagger}}
       {\mathcal{E}_n(\bm{\lambda})-\mathcal{E}_{n-1}(\bm{\lambda})}\,
  \phi_{n,n}(y\,;\bm{\lambda})\,,
  \label{rat_CS_phi=Adphi}
\end{align}
and from \eqref{rat_CS_H_rec} we have
\begin{equation}
  \phi_{n,n}(y\,;\bm{\lambda})\propto\phi_0(y\,;\bm{\lambda}+n\bm{\delta}).
  \label{rat_CS_phi=phi}
\end{equation}
It should be emphasized that all the operators $\mathcal{A}$ and
$\mathcal{A}^\dagger$ in the above formulas are explicitly known thanks
to the shape-invariance.
The latter formula \eqref{rat_CS_phi=Adphi} with \eqref{rat_CS_phi=phi}
can be understood as the Rodrigues-type formula.
The relation \eqref{rat_CS_E_rec} means that
$\{\mathcal{E}_n(\bm{\lambda})\}_{n\geq 0}$ is calculable from
$\mathcal{E}_1(\bm{\lambda})$, namely the spectrum is determined by the
shape invariance.
In the present case we obtain
\begin{equation}
  \mathcal{E}_n=n.
  \label{ratA_CS_E}
\end{equation}

\begin{figure}
  \centering
  \includegraphics*[scale=.7]{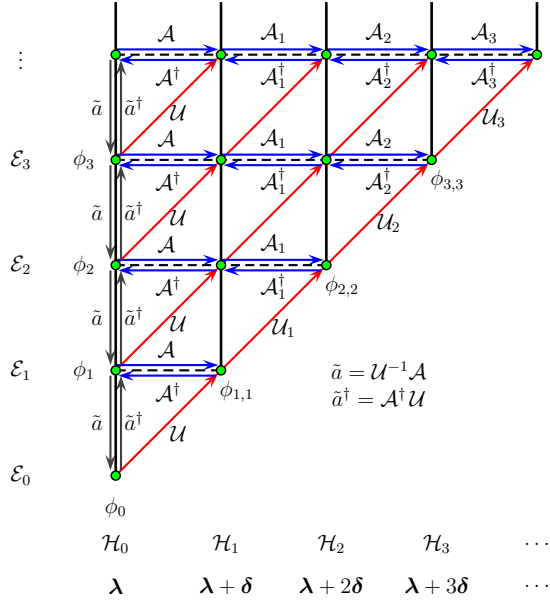}
  \caption{A schematic diagram of the energy levels and the associated
  Hamiltonian systems  together with the definition of the $\mathcal{A}$
  and $\mathcal{A}^\dagger$ operators and the `creation'
  ($\tilde{a}^{\dagger}$) and `annihilation' ($\tilde{a}$) operators.
  The parameter set is indicated below each Hamiltonian.}
\end{figure}

As seen above, the operators $\mathcal{A}$ and $\mathcal{A}^\dagger$ act
isospectrally, that is horizontally. On the other hand, the annihilation
and creation operators map from one eigenstate to another, {\em i.e.\/} 
vertically, of a given Hamiltonian.
In order to define the annihilation and creation operators, let us
introduce normalized basis $\{\hat{\phi}_{s,n}\}_{n\ge s}$ for each
Hamiltonian $\mathcal{H}_s$ and unitary operators
$\mathcal{U}_s=\mathcal{U}_s(\bm{\lambda})$ mapping
the $s$-th orthonormal basis $\{\hat{\phi}_{s,n}\}_{n\ge s}$ to the
$(s+1)$-th $\{\hat{\phi}_{s+1,n}\}_{n\ge s+1}$ (see Fig.1 and for
example \cite{SVZ93,KD02}):
\begin{equation}
  \mathcal{U}_s(\bm{\lambda})\hat{\phi}_{s,n}(y\,;\bm{\lambda})
  =\hat{\phi}_{s+1,n+1}(y\,;\bm{\lambda}),\quad
  \mathcal{U}_s(\bm{\lambda})^\dagger \hat{\phi}_{s+1,n+1}(y\,;\bm{\lambda})
  =\hat{\phi}_{s,n}(y\,;\bm{\lambda}).
\end{equation}
We denote $\mathcal{U}_0=\mathcal{U}$.
Roughly speaking $\mathcal{U}$ increases the parameters from
$\bm{\lambda}$ to $\bm{\lambda}+\bm{\delta}$.
Then an annihilation $\tilde{a}$ and a creation operator
$\tilde{a}^\dagger$ for the Hamiltonian $\mathcal{H}$ are introduced
as follows:
\begin{equation}
  \tilde{a}=\tilde{a}(y\,;\bm{\lambda})\eqdef
  \mathcal{U}^{\dagger}(\bm{\lambda})\mathcal{A}(y\,;\bm{\lambda}),
  \quad
  \tilde{a}^{\dagger}=\tilde{a}(y\,;\bm{\lambda})^{\dagger}\eqdef
  \mathcal{A}(y\,;\bm{\lambda})^{\dagger}\mathcal{U}(\bm{\lambda}).
\end{equation}
It is straightforward to derive
\begin{gather}
  \mathcal{H}(y\,;\bm{\lambda})
  =\tilde{a}(y\,;\bm{\lambda})^{\dagger}\tilde{a}(y\,;\bm{\lambda}),\\
  \bigl[\tilde{a}(y\,;\bm{\lambda}),\tilde{a}(y\,;\bm{\lambda})^{\dagger}\bigr]
  \hat{\phi}_n(y\,;\lambda)
  =({\cal E}_{n+1}(\bm{\lambda})-{\cal E}_n(\bm{\lambda}))
  \hat{\phi}_n(y\,;\lambda).
\end{gather}
In the present case $U$ is an identity map and we recover the well-known
result.
This scheme is illustrated in Figure 1.

The above Rodrigues-type formula
\eqref{rat_CS_phi=Adphi}--\eqref{rat_CS_phi=phi}
gives $\phi_n(y)\propto H_n(y)\phi_0(y)$.
This can be also understood in the following manner.
By similarity transformation in terms of the ground state wavefunction,
let us define $\tilde{\mathcal{H}}$,
\begin{align}
  \tilde{\mathcal{H}}&=\phi_0(y\,;\bm{\lambda})^{-1}\circ
  \mathcal{H}\circ\phi_0(y\,;\bm{\lambda})
  =-\frac12\frac{d^2}{dy^2}
  -\frac{d\,\mathcal{W}(y\,;\bm{\lambda})}{dy}\frac{d}{dy}\,,
  \label{rat_CS_tcalH}\\
  &=BC,\quad
  B=-i\bigl(\tfrac{d}{dy}+2\tfrac{d\,\mathcal{W}(y\,;\bm{\lambda})}{dy}\bigr),
  \quad C=-\tfrac{i}{2}\tfrac{d}{dy},
  \label{rat_CS_BC}
\end{align}
and consider higher eigenfunctions in a product form
$\phi_n(y\,;\bm{\lambda})=P_n(y\,;\bm{\lambda})\phi_0(y\,;\bm{\lambda})$,
where $P_n(y\,;\bm{\lambda})$ satisfies
\begin{equation}
  \tilde{\mathcal{H}}(y\,;\bm{\lambda})P_n(y\,;\bm{\lambda})
  =\mathcal{E}_n(\bm{\lambda})P_n(y\,;\bm{\lambda}).
  \label{cHP=cEP}
\end{equation}
In the present case we have
\begin{equation}
  \tilde{\mathcal{H}}=-\frac12\frac{d^2}{dy^2}+y\frac{d}{dy}\,.
\end{equation}
Since the Hermite polynomial satisfies
\begin{equation}
  \Bigl(\frac{d^2}{dy^2}-2y\frac{d}{dy}+2n\Bigr)H_n(y)=0,
\end{equation}
we obtain
\begin{equation}
  P_n(y)\propto H_n(y),\quad \mathcal{E}_n=n.
\end{equation}

The energy of $H$ is
\begin{equation}
  E_n=\hbar\omega n\,.
  \label{ratA_CS_Eorg}
\end{equation}

\subsection{Ruijsenaars-Schneider-van Diejen Systems}

The Hamiltonian is \eqref{H_RS} with the potential
\eqref{ratA_RS_v}--\eqref{ratA_RS_w}.

\subsubsection{Equilibrium positions of $n$-particle classical systems}

Let us consider a polynomial whose zeros give the equilibrium positions,
$f(y)=\prod_{j=1}^n(y-\sqrt{\frac{m\omega_1}{\bar{g}}}\,\bar{q}_j)$.
Then \eqref{equiveqBA} can be converted to a functional equation
for $f(y)$. We can show that the solutions of this functional
equation satisfy the three-term recurrence which agrees with that of the
continuous Hahn polynomials of specific parameters.
The result is \cite{OS2}
\begin{equation}
  \prod_{j=1}^n\Bigl(y-\sqrt{\frac{m\omega_1}{\bar{g}}}\,\bar{q}_j\Bigr)
  =p_n^{\rm monic}\Bigl(\sqrt{\frac{mc^2}{\omega_1\bar{g}}}\,y\,;
  \frac{mc^2}{\omega_1\bar{g}},\frac{mc^2}{\omega_2\bar{g}},
  \frac{mc^2}{\omega_1\bar{g}},\frac{mc^2}{\omega_2\bar{g}}\Bigr)\,,
\end{equation}
where $p_n(y;a_1,a_2,b_1,b_2)=\frac{1}{n!}(n+a_1+a_2+b_1+b_2-1)_n\,
p_n^{\rm monic}(y;a_1,a_2,b_1,b_2)$ is the continuous Hahn polynomial
\cite{KS96}.

\subsubsection{Eigenfunctions of single-particle quantum mechanics}
\label{1particle_ratA_RS}

Let us consider single-particle case ($n=1$).
The potential $V_1(q)$ is simply $V_1(q)=w(q_1)$. Let us write $x=q_1$.
The Hamiltonian \eqref{H_RS} becomes
\begin{equation}
  H=\frac{mc^2}{2}\Bigl(
  \sqrt{w(x)}\,e^{-i\frac{\hbar}{mc}\frac{d}{dx}}\sqrt{w(x)^*}
  +\sqrt{w(x)^*}\,e^{i\frac{\hbar}{mc}\frac{d}{dx}}\sqrt{w(x)}
  -w(x)-w(x)^*\Bigr)\,.
  \label{rat_RS_H}
\end{equation}
By introducing a dimensionless variable\footnote{
Since $q_j$ and $p_j$ are same in both hand sides of \eqref{RSCS},
$y$ here and $y$ in \eqref{rat_CS_y} are different:
$\frac{y\text{ in \eqref{rat_CS_y}}}{y\text{ in \eqref{rat_RS_y}}}
=\sqrt{\frac{\hbar\omega}{mc^2}}$.
In order to take $c\rightarrow\infty$ limit we should rescale $y$ here
$c$-dependently.
}
$y$ and a rescaled potential $V(y)$,
\begin{gather}
  y=\frac{mc}{\hbar}x\,,
  \label{rat_RS_y}\\
  V(y)=V\bigl(y\,;(a_1,a_2)\bigr)=(a_1+iy)(a_2+iy)\,,
  \label{ratA_RS_V}
\end{gather}
$w(x)$ and $H$ are expressed as
\begin{gather}
  w(x)=\frac{\hbar\omega_1}{mc^2}\frac{\hbar\omega_2}{mc^2}
  V\Bigl(y\,;\bigl(\frac{mc^2}{\hbar\omega_1},\frac{mc^2}{\hbar\omega_2}
  \bigr)\Bigr)\,,\\
  H=mc^2\frac{\hbar\omega_1}{mc^2}\frac{\hbar\omega_2}{mc^2}\mathcal{H}\,.
\end{gather}
Here $\mathcal{H}$ is defined by
\begin{equation}
  \mathcal{H}=\frac12\Bigl(
  \sqrt{V(y)}\,e^{-i\frac{d}{dy}}\sqrt{V(y)^*}
  +\sqrt{V(y)^*}\,e^{i\frac{d}{dy}}\sqrt{V(y)}
  -V(y)-V(y)^*\Bigr)\,,
  \label{rat_RS_calH}
\end{equation}
where $V(y)$ is $V(y;\bm{\lambda})$ \eqref{ratA_RS_V} with
$\bm{\lambda}=(a_1,a_2)=
(\frac{mc^2}{\hbar\omega_1},\frac{mc^2}{\hbar\omega_2})$.
In the following we will consider arbitrary positive parameters
$a_1$ and $a_2$.
Instead of $H\phi_n=E_n\phi_n$, let us consider a rescaled equation
$\mathcal{H}\phi_n(y)=\mathcal{E}_n\phi_n(y)$ $(n=0,1,2,\ldots)$,
where energies are related as
$E_n=\frac{\hbar^2\omega_1\omega_2}{mc^2}\mathcal{E}_n$.

The Hamiltonian $\mathcal{H}$ is factorizable:
\begin{align}
  \mathcal{H}&=\mathcal{H}(y\,;\bm{\lambda})
  =\mathcal{A}(y\,;\bm{\lambda})^{\dagger}\mathcal{A}(y\,;\bm{\lambda})\,,
  \label{rat_RS_calH=AA}\\
  \mathcal{A}&=\mathcal{A}(y\,;\bm{\lambda})\eqdef\frac{1}{\sqrt{2}}\Bigl(
  e^{-\frac{i}{2}\frac{d}{dy}}\sqrt{V(y\,;\bm{\lambda})^*}
  -e^{\frac{i}{2}\frac{d}{dy}}\sqrt{V(y\,;\bm{\lambda})}\,\Bigr)\,,\\
  \mathcal{A}^{\dagger}&=\mathcal{A}(y\,;\bm{\lambda})^{\dagger}
  \eqdef\frac{1}{\sqrt{2}}\Bigl(
  \sqrt{V(y\,;\bm{\lambda})}\,e^{-\frac{i}{2}\frac{d}{dy}}
  -\sqrt{V(y\,;\bm{\lambda})^*}\,e^{\frac{i}{2}\frac{d}{dy}}\Bigr)\,.
  \label{rat_RS_Ad}
\end{align}
The ground state of $\mathcal{H}$ is annihilated by $\mathcal{A}$,
\begin{equation}
  \phi_0(y\,;\bm{\lambda})\propto
  \bigl|\Gamma(a_1+iy)\Gamma(a_2+iy)\bigr|\,,\quad
  \mathcal{E}_0(\bm{\lambda})=0.
\end{equation}
It is easy to verify that the Hamiltonian $\mathcal{H}$ is shape
invariant \eqref{rat_CS_shapeinv} with
\begin{equation}
  \bm{\delta}=(\tfrac12,\tfrac12)\,,\quad
  \mathcal{E}_1(\bm{\lambda})=a_1+a_2\,.
  \label{ratA_RS_E1}
\end{equation}
Like in \S\ref{1particle_ratA_CS}, let us define
$\mathcal{A}_s$, $\mathcal{H}_s$ and $\phi_{s,n}$ ($n\geq s\geq 0$)
by \eqref{rat_CS_A_rec}--\eqref{rat_CS_phi_rec}.
Then for $n\geq s\geq 0$ we obtain
\eqref{rat_CS_As}--\eqref{rat_CS_Adphi}
and \eqref{rat_CS_phi=Aphi}--\eqref{rat_CS_phi=phi}.
{}From \eqref{rat_CS_E_rec} and \eqref{ratA_RS_E1} we get
\begin{equation}
  \mathcal{E}_n(\bm{\lambda})=\tfrac12n(n+2a_1+2a_2-1).
  \label{ratA_RS_E}
\end{equation}

By similarity transformation in terms of the ground state wavefunction,
we define $\tilde{\mathcal{H}}$,
\begin{align}
  \tilde{\mathcal{H}}&=\phi_0(y\,;\bm{\lambda})^{-1}\circ
  \mathcal{H}\circ\phi_0(y\,;\bm{\lambda})
  =\frac12\Bigl(V(y)e^{-i\frac{d}{dy}}+V(y)^*e^{i\frac{d}{dy}}
  -V(y)-V(y)^*\Bigr),
  \label{rat_RS_tcalH}\\
  &=BC,\quad
  B=-i\bigl(V(y)e^{-\frac{i}{2}\frac{d}{dy}}
  -V(y)^*e^{\frac{i}{2}\frac{d}{dy}}\bigr),\quad
  C=\tfrac{i}{2}\bigl(e^{-\frac{i}{2}\frac{d}{dy}}
  -e^{\frac{i}{2}\frac{d}{dy}}\bigr),
  \label{rat_RS_BC}
\end{align}
and consider
$\phi_n(y\,;\bm{\lambda})=P_n(y\,;\bm{\lambda})\phi_0(y\,;\bm{\lambda})$,
where $P_n(y\,;\bm{\lambda})$ satisfies \eqref{cHP=cEP}.
This means that $P_n(y\,;\bm{\lambda})$ is a special case of the
continuous Hahn polynomial
\begin{equation}
  P_n(y;\bm{\lambda})\propto p_n(y\,;a_1,a_2,a_1,a_2),\quad
  \mathcal{E}_n(\bm{\lambda})=\tfrac12n(n+2a_1+2a_2-1).
\end{equation}

The energy of $H$ is
\begin{equation}
  E_n=\hbar(\omega_1+\omega_2)n+\frac{\hbar^2\omega_1\omega_2}{2mc^2}n(n-1).
\end{equation}
In the $c\rightarrow\infty$ limit we have
$\displaystyle\lim_{c\rightarrow\infty}E_n=\hbar(\omega_1+\omega_2)n$
and this is consistent with \eqref{RSCS}, \eqref{ratA_RSCS_para} and
\eqref{ratA_CS_Eorg}.

\section{Rational $BC$ Types}

In this section we consider CS and RSvD systems with rational $BC$ type
potentials.
Relevant polynomials are the Laguerre polynomial and the Wilson polynomial.

\subsection{Calogero Systems}

The Hamiltonian is \eqref{H_CS} with the potential
\eqref{ratBC_CS_V}--\eqref{ratBC_CS_W}.

\subsubsection{Equilibrium positions of $n$-particle classical systems}

Let us consider a polynomial whose zeros give the equilibrium positions,
$f(y)=\prod_{j=1}^n(y^2-\frac{m\omega}{\bar{g}_M}\,\bar{q}_j^2)$.
Then \eqref{CS_equiveqW} can be converted to a differential equation
for $f(y)$, which is the differential equation for the Laguerre
polynomial. The result is
\begin{equation}
  \prod_{j=1}^n\Bigl(y^2-\frac{m\omega}{\bar{g}_M}\,\bar{q}_j^2\Bigr)
  =L_n^{(\alpha)\,{\rm monic}}(y^2)\,,\quad
  \alpha=\frac{\bar{g}_S+\bar{g}_L}{\bar{g}_M}-1\,,
\end{equation}
where $L_n^{(\alpha)}(y^2)=\frac{(-1)^n}{n!}L_n^{(\alpha)\,{\rm monic}}(y^2)$
is the Laguerre polynomial \cite{KS96}.

\subsubsection{Eigenfunctions of single-particle quantum mechanics}
\label{1particle_ratBC_CS}

Let us consider the single-particle case ($n=1$) and write $x=q_1$ and
$g=g_S+g_L$. The Hamiltonian \eqref{H_CS} describes the harmonic oscillator
with a centrifugal barrier and a constant energy shift
\begin{equation}
  H=-\frac{\hbar^2}{2m}\frac{d^2}{dx^2}+\frac12m\omega^2x^2
  +\frac{\hbar^2}{2m}\frac{g(g-1)}{x^2}-\hbar\omega\Bigl(g+\frac12\Bigr)\,.
\end{equation}
By introducing a dimensionless variable $y$ \eqref{rat_CS_y},
$H$ can be written as
\begin{equation}
  H=\hbar\omega\mathcal{H}\,,\quad
  \mathcal{H}=-\frac12\frac{d^2}{dy^2}+\frac12y^2
  +\frac{g(g-1)}{2y^2}-g-\frac12\,.
\end{equation}
This $\mathcal{H}$ has a parameter $g$ and we will write $\bm{\lambda}=g$.
Instead of $H\phi_n=E_n\phi_n$, let us consider a rescaled equation
$\mathcal{H}\phi_n(y)=\mathcal{E}_n\phi_n(y)$ ($n=0,1,2,\ldots$),
where energies are related as $E_n=\hbar\omega\mathcal{E}_n$.

Like in \S\ref{1particle_ratA_CS}, $\mathcal{H}$ is factorizable
\eqref{rat_CS_calH=AA}--\eqref{rat_CS_Ad}, where the prepotential with
parameter $\bm{\lambda}=g$ is
\begin{equation}
  \mathcal{W}(y\,;\bm{\lambda})=-\tfrac12y^2+g\log y.
  \label{ratBC_CS_cW}
\end{equation}
The ground state wavefunction of $\mathcal{H}$ is \eqref{rat_CS_phi0},
\begin{equation}
  \phi_0(y\,;\bm{\lambda})\propto y^ge^{-\frac12y^2},\quad
  \mathcal{E}_0(\bm{\lambda})=0.
  \label{ratBC_CS_phi0}
\end{equation}
The Hamiltonian $\mathcal{H}$ is shape invariant \eqref{rat_CS_shapeinv}
with
\begin{equation}
  \bm{\delta}=1,\quad
  \mathcal{E}_1(\bm{\lambda})=2.
  \label{ratBC_CS_E1}
\end{equation}
Let us define
$\mathcal{A}_s$, $\mathcal{H}_s$ and $\phi_{s,n}$ ($n\geq s\geq 0$)
by \eqref{rat_CS_A_rec}--\eqref{rat_CS_phi_rec}.
Then for $n\geq s\geq 0$ we arrive at the consequence of the
shape-invariance \eqref{rat_CS_As}--\eqref{rat_CS_Adphi}
and \eqref{rat_CS_phi=Aphi}--\eqref{rat_CS_phi=phi}.
{}From \eqref{rat_CS_E_rec} and \eqref{ratBC_CS_E1} we find the energy
spectrum 
\begin{equation}
  \mathcal{E}_n(\bm{\lambda})=2n.
  \label{ratBC_CS_E}
\end{equation}

The Rodrigues-type formula \eqref{rat_CS_phi=Adphi}--\eqref{rat_CS_phi=phi}
gives $\phi_n(y\,;g)\propto L_n^{(g-\frac12)}(y^2)\phi_0(y\,;g)$.
This can be also understood as \eqref{rat_CS_tcalH}--\eqref{cHP=cEP},
\begin{equation}
  \tilde{\mathcal{H}}
  =-\frac12\frac{d^2}{dy^2}+\bigl(y-\frac{g}{y}\bigr)\frac{d}{dy}
  =2\Bigl(-\eta\frac{d^2}{d\eta^2}
  +(\eta-g-\tfrac12)\frac{d}{d\eta}\Bigr)\,,\quad(\eta=y^2).
\end{equation}
Since the Laguerre polynomial satisfies
\begin{equation}
  \Bigl(\eta\frac{d^2}{d\eta^2}+(\alpha+1-\eta)\frac{d}{d\eta}+n\Bigr)
  L_n^{(\alpha)}(\eta)=0,
\end{equation}
we obtain
\begin{equation}
  P_n(y\,;g)\propto L_n^{(g-\frac12)}(y^2),\quad
  \mathcal{E}_n(g)=2n.
\end{equation}

The energy of $H$ is
\begin{equation}
  E_n(g)=\hbar\omega 2n.
  \label{ratBC_CS_Eorg}
\end{equation}

\noindent
\underline{Remark}:
For the given potential $V_{\rm CS}(x)$, the prepotential $W(x)$ is
obtained by solving the Riccati equation $V_{\rm CS}(x)
=\frac{1}{2m}(W'(x)^2+\hbar W^{\prime\prime}(x))+\text{constant}$.
The above prepotential \eqref{ratBC_CS_cW} is one solution.
Since the Hamiltonian(except for the constant term) contains $g$ as the
combination $g(g-1)$, we have another solution $\check{\mathcal{W}}$,
\begin{equation}
  \check{\mathcal{W}}(y\,;g)=\mathcal{W}(y\,;1-g),
\end{equation}
and the corresponding Hamiltonian $\check{\mathcal{H}}$ is
\begin{equation}
  \check{\mathcal{H}}(y\,;g)=\check{\mathcal{A}}(y\,;g)^{\dagger}
  \check{\mathcal{A}}(y\,;g)=\mathcal{H}(y\,;g)+2g-1.
\end{equation}
The ground state wavefunction of $\check{\mathcal{H}}$ (the state
annihilated by $\check{\mathcal{A}}(y\,;g)$) is
\begin{equation}
  \check{\phi}_0(y\,;g)\propto e^{\check{\mathcal{W}}(y\,;g)}
  =y^{1-g}e^{-\frac12y^2},\quad
  \check{\mathcal{E}}_0(g)=0,
\end{equation}
which is square integrable for $g<\frac32$.
Note that $\phi_0(y\,;g)$ \eqref{ratBC_CS_phi0} is square integrable 
for $g>-\frac12$.
These two `ground' states define two sectors of this system.
Usually we consider only one of them. 
Note that $\mathcal{A}(y\,;g)\check{\phi}_0(y\,;g)$ is square integrable
for $g\leq\frac12$.
{}From the Rodrigues-type formula and the recurrence relation of the
energy, we obtain
\begin{equation}
  \check{\phi}_n(y\,;g)\propto L_n^{(\frac12-g)}(y^2)\check{\phi}_0(y\,;g),
  \quad \check{\mathcal{E}}_n(g)=2n.
  \label{ratBC_CS_checkphi}
\end{equation}
The corresponding energy of $H$ is
\begin{equation}
  \check{E}_n(g)=\hbar\omega(2n+1-2g).
  \label{ratBC_CS_E'org}
\end{equation}
Therefore, for $g<\frac32$ ($g\neq\frac12$), we have another sector of
the system,  \eqref{ratBC_CS_checkphi} with \eqref{ratBC_CS_E'org}.
The order of $E_n(g)$ \eqref{ratBC_CS_Eorg} and $\check{E}_n(g)$
\eqref{ratBC_CS_E'org} is
\begin{align}
  E_0(g)<\check{E}_0(g)<E_1(g)<\check{E}_1(g)<E_2(g)<\check{E}_2(g)<\cdots
  \quad&\text{ for } -\tfrac12<g<\tfrac12\,,\\
  \check{E}_0(g)<E_0(g)<\check{E}_1(g)<E_1(g)<\check{E}_2(g)<E_2(g)<\cdots
  \quad&\text{ for } \quad\ \tfrac12<g<\tfrac32\,.
\end{align}
Thus the lowest energy state of $H$ is $\phi_0(y\,;g)$ for
$g\geq\frac32$ or $-\frac12<g\leq\frac12$ (which cover all values of
$g(g-1)\geq-\frac14$), and $\check{\phi}_0(y\,;g)$
for $\frac12<g<\frac32$.
In the $g\rightarrow 0$(or 1) limit, both sectors contribute and
these eigenfunctions reduce to those in \S\ref{1particle_ratA_CS} due to
the identities,
\begin{gather}
  \phi_n(y\,;g\rightarrow 0)\propto
  L_n^{(-\frac12)}(y^2)\,e^{-\frac12y^2}\propto H_{2n}(y)\,e^{-\frac12y^2},\\
  \check{\phi}_n(y\,;g\rightarrow 0)\propto
  L_n^{(\frac12)}(y^2)\,y\,e^{-\frac12y^2}\propto H_{2n+1}(y)\,e^{-\frac12y^2}.
\end{gather}

\subsection{Ruijsenaars-Schneider-van Diejen Systems}

The Hamiltonian is \eqref{H_RS} with the potential
\eqref{ratBC_RS_v}--\eqref{ratBC_RS_w}.

\subsubsection{Equilibrium positions of $n$-particle classical systems}

Let us consider a polynomial whose zeros give the equilibrium positions,
$f(y)=\prod_{j=1}^n(y^2-\frac{m\omega_1}{\bar{g}_0}\,\bar{q}_j^2)$.
Then \eqref{equiveqBA} can be converted to a functional equation
for $f(y)$. We can show that the solutions of this functional
equation satisfy the three-term recurrence which agrees with that of the
Wilson polynomials. The result is \cite{OS2}
\begin{equation}
  \prod_{j=1}^n\Bigl(y^2-\frac{m\omega_1}{\bar{g}_0}\,\bar{q}_j^2\Bigr)
  =W_n^{\rm monic}\Bigl(\frac{mc^2}{\omega_1\bar{g}_0}\,y^2;
  \frac{mc^2}{\omega_1\bar{g}_0},\frac{mc^2}{\omega_2\bar{g}_0},
  \frac{\bar{g}_1}{\bar{g}_0},\frac{\bar{g}_2}{\bar{g}_0}\Bigr)\,,
\end{equation}
where $W_n(y^2;a_1,a_2,a_3,a_4)=(-1)^n(n+a_1+a_2+a_3+a_4-1)_n\,
W_n^{\rm monic}(y^2;a_1,a_2,a_3,a_4)$ is the Wilson polynomial
\cite{KS96}.

\subsubsection{Eigenfunctions of single-particle quantum mechanics}
\label{1particle_ratBC_RS}

Let us consider the single-particle case ($n=1$).
The potential $V_1(q)$ is $V_1(q)=w(q_1)$. Let us write $x=q_1$.
The Hamiltonian \eqref{H_RS} becomes \eqref{rat_RS_H} with $w(x)$ in
\eqref{ratBC_RS_w}.
By introducing a dimensionless variable $y$ \eqref{rat_RS_y} and
a rescaled potential $V(y)$,
\begin{equation}
  V(y)=V\bigl(y\,;(a_1,a_2,a_3,a_4)\bigr)
  =\frac{(a_1+iy)(a_2+iy)(a_3+iy)(a_4+iy)}{2iy(2iy+1)}\,,
  \label{ratBC_RS_V}
\end{equation}
$w(x)$ and $H$ are expressed as
\begin{gather}
  w(x)=4\frac{\hbar\omega_1}{mc^2}\frac{\hbar\omega_2}{mc^2}
  V\Bigl(y\,;\bigl(\frac{mc^2}{\hbar\omega_1},\frac{mc^2}{\hbar\omega_2},
  g_1,g_2\bigr)\Bigr)\,,\\
    H=4mc^2\frac{\hbar\omega_1}{mc^2}\frac{\hbar\omega_2}{mc^2}\mathcal{H}\,.
\end{gather}
Here rescaled Hamiltonian $\mathcal{H}$ is defined by \eqref{rat_RS_calH},
where $V(y)$ is $V(y;\bm{\lambda})$ \eqref{ratBC_RS_V} with
$\bm{\lambda}=(a_1,a_2,a_3,a_4)=
(\frac{mc^2}{\hbar\omega_1},\frac{mc^2}{\hbar\omega_2},g_1,g_2)$.
In the following we will consider arbitrary positive parameters
$a_1,a_2,a_3,a_4$.
Instead of $H\phi_n=E_n\phi_n$, let us consider the rescaled equation
$\mathcal{H}\phi_n(y)=\mathcal{E}_n\phi_n(y)$ $(n=0,1,2,\ldots)$,
where energies are related as
$E_n=\frac{4\hbar^2\omega_1\omega_2}{mc^2}\mathcal{E}_n$.

Like in \S\ref{1particle_ratA_RS}, $\mathcal{H}$ is factorizable
\eqref{rat_RS_calH=AA}--\eqref{rat_RS_Ad}.
The ground state of $\mathcal{H}$ is annihilated by $\mathcal{A}$,
\begin{equation}
  \phi_0(y\,;\lambda)\propto \biggl|
  \frac{\Gamma(a_1+iy)\Gamma(a_2+iy)\Gamma(a_3+iy)\Gamma(a_4+iy)}
  {\Gamma(2iy)}\biggr|\,.
\end{equation}
The Hamiltonian $\mathcal{H}$ is shape invariant \eqref{rat_CS_shapeinv}
with 
\begin{equation}
  \bm{\delta}=(\tfrac12,\tfrac12,\tfrac12,\tfrac12),\quad
  \mathcal{E}_1(\bm{\lambda})=\tfrac12(a_1+a_2+a_3+a_4).
  \label{ratBC_RS_E1}
\end{equation}
The third and fourth components of $\bm{\delta}$ are consistent with
$\bm{\delta}$ in \eqref{ratBC_CS_E1}
because of \eqref{RSCS} and \eqref{ratBC_RSCS_para}.
Let us define
$\mathcal{A}_s$, $\mathcal{H}_s$ and $\phi_{s,n}$ ($n\geq s\geq 0$)
by \eqref{rat_CS_A_rec}--\eqref{rat_CS_phi_rec}.
Then for $n\geq s\geq 0$ we obtain the consequences of the
shape-invariance \eqref{rat_CS_As}--\eqref{rat_CS_Adphi}
and \eqref{rat_CS_phi=Aphi}--\eqref{rat_CS_phi=phi}.
{}From \eqref{rat_CS_E_rec} and \eqref{ratBC_RS_E1} we obtain the 
energy spectrum
\begin{equation}
  \mathcal{E}_n(\bm{\lambda})=\frac12n(n+a_1+a_2+a_3+a_4-1).
\end{equation}
By similarity transformation in terms of the ground state wavefunction,
\eqref{rat_RS_tcalH}--\eqref{rat_RS_BC} and \eqref{cHP=cEP} imply that
$P_n(y\,;\bm{\lambda})$ is the Wilson polynomial
\begin{equation}
  P_n(y\,;\bm{\lambda})\propto W_n(y^2;\bm{\lambda}),\quad
  \mathcal{E}_n(\bm{\lambda})=\frac12n(n+a_1+a_2+a_3+a_4-1).
\end{equation}

The energy of $H$ is
\begin{equation}
  E_n=\hbar(\omega_1+\omega_2)2n
  +\frac{\hbar^2\omega_1\omega_2}{mc^2}2n(n+g_1+g_2-1).
\end{equation}
In the $c\rightarrow\infty$ limit we have
$\displaystyle\lim_{c\rightarrow\infty}E_n=\hbar(\omega_1+\omega_2)2n$
and this is consistent with \eqref{RSCS}, \eqref{ratBC_RSCS_para} and
\eqref{ratBC_CS_Eorg}.

\section{Trigonometric $A$ Types}

In this section we consider the CS and RSvD systems with the
trigonometric $A$ type potentials.
The single-particle quantum mechanics is free theory and cosine (or sine)
functions are the eigenfunctions.

\subsection{Sutherland Systems}

The Hamiltonian is \eqref{H_CS} with the potential
\eqref{trigA_CS_V}--\eqref{trigA_CS_W}.

\subsubsection{Equilibrium positions of $n$-particle classical systems}
\label{eqiuv_trigA_CS}

The equation for the equilibrium positions \eqref{CS_equiveqW} is
easily solved,
\begin{equation}
  \frac{\pi}{L}\bar{q}_j=\frac{\pi}{n}(n+1-j)+\alpha\qquad
  (j=1,\ldots,n),
  \label{equivpo_trigA_CS}
\end{equation}
where $\alpha$ is an arbitrary real number which is a consequence of
the translational invariance.
The rescaled equilibrium positions $\frac{\pi}{L}\bar{q}_j$ are zeros of
$\cos n(\theta-\alpha')$ ($\alpha'=\alpha-\frac{\pi}{2n}$), which is
equal to
\begin{equation}
  T_n\bigl(\cos(\theta-\alpha')\bigr).
\end{equation}
Here $T_n(\cos\varphi)=\cos n\varphi$ is the Chebyshev polynomial of
the first kind \cite{KS96}.

\subsubsection{Eigenfunctions of single-particle quantum mechanics}
\label{1particle_trigA_CS}

Let us consider the single-particle case ($n=1$) and write $x=q_1$.
We impose the periodic boundary condition on the wave function $\phi(x)$,
$\phi(x+L)=\phi(x)$. The Hamiltonian \eqref{H_RS} is a free one
\begin{equation}
  H=-\frac{\hbar^2}{2m}\frac{d^2}{dx^2}.
\end{equation}
By introducing a dimensionless variable $y$,
\begin{equation}
  y=\frac{\pi}{L}x,
  \label{trig_CS_y}
\end{equation}
$H$ can be written as
\begin{equation}
  H=\frac{\hbar^2\pi^2}{mL^2}\mathcal{H}\,,\quad
  \mathcal{H}=-\frac12\frac{d^2}{dy^2}\,.
  \label{trigA_CS_cH}
\end{equation}
Moreover in terms of another dimensionless variable $z$,
\begin{equation}
  z=e^{2iy}=e^{2\pi i\frac{x}{L}},
  \label{trig_CS_z}
\end{equation}
$\mathcal{H}$ becomes
\begin{equation}
  \mathcal{H}=2D_z^2\,,\quad D_z\eqdef z\frac{d}{dz}\,.
\end{equation}
The eigenfunctions of $\mathcal{H}$ (with periodic boundary condition in $x$)
are easily obtained: $z^n$ ($n\in\mathbb{Z}$).
Except for the ground state, eigenstates are doubly degenerated,
\begin{gather}
  \phi_0(z)\propto1,\quad
  \phi_n(z)\propto c_1z^n+c_2z^{-n}\propto\cos n(2y-\alpha')
  =T_n(\cos(2y-\alpha'))\quad(n\geq 1),
  \label{trigA_CS_phi}\\
  \mathcal{E}_n=2n^2\quad(n\geq 0),
\end{gather}
where $c_1,c_2,\alpha'$ are arbitrary numbers.
The variable $2y$ should be identified with $\theta$ in
\S\ref{eqiuv_trigA_CS} (see \S\ref{1particle_trigBC_CS}).
The energy spectrum of $H$ is
\begin{equation}
  E_n=\frac{\hbar^2\pi^2}{mL^2}2n^2\,.
  \label{trigA_CS_Eorg}
\end{equation}

\subsection{Ruijsenaars-Schneider Systems}

The Hamiltonian is \eqref{H_RS} with the potential
\eqref{trigA_RS_v}--\eqref{trigA_RS_w}.

\subsubsection{Equilibrium positions of $n$-particle classical systems}

The equation for the equilibrium positions \eqref{RS_equiveq} is
easily solved and the equilibrium positions are the same as those given
in \S\ref{eqiuv_trigA_CS}, \eqref{equivpo_trigA_CS}.

\subsubsection{Eigenfunctions of single-particle quantum mechanics}

Let us consider the single-particle case ($n=1$).
The potential $V_1(q)$ is trivial $V_1(q)=w(q_1)=1$. Let us write $x=q_1$.
We impose the periodic boundary condition $\phi(x+L)=\phi(x)$, too.
The Hamiltonian \eqref{H_RS} becomes
\begin{equation}
  H=\frac{mc^2}{2}\bigl(e^{-i\frac{\hbar}{mc}\frac{d}{dx}}
  +e^{i\frac{\hbar}{mc}\frac{d}{dx}}-2\bigr)\,.
  \label{trigA_RS_H}
\end{equation}
By introducing a dimensionless variable $y$ \eqref{trig_CS_y} and
$z$ \eqref{trig_CS_z}, $H$ can be written as
\begin{equation}
  H=mc^2\mathcal{H}\,,\quad
  \mathcal{H}=\tfrac12\bigl(e^{-i\frac{d}{dy}}+e^{i\frac{d}{dy}}-2\bigr)
  =\tfrac12\bigl(q^{-D_z}+q^{D_z}-2\bigr),
\end{equation}
where we have introduced a dimensionless parameter\footnote{
Here we adopt the standard notation for the modulus $q$. There should not
be any confusion with the coordinate $q$.
}
$q$,
\begin{equation}
  q=e^{-\frac{2\pi\hbar}{mcL}}, \quad 0<q<1.
  \label{q}
\end{equation}
The operator $q^{D_z}$ causes a $q$-shift, $q^{D_z}f(z)=f(qz)$.
Again this is a free theory and the eigenfunctions of $\mathcal{H}$ 
(with periodic boundary condition in $x$)
are easily obtained: $z^n$ ($n\in\mathbb{Z}$).
Except for the ground state, eigenstates are doubly degenerate:
\eqref{trigA_CS_phi} and
\begin{equation}
  \mathcal{E}_n=2\sinh^2\frac{\hbar\pi n}{mcL}\quad(n\geq 0).
\end{equation}

The energy spectrum of $H$ is
\begin{equation}
  E_n=2mc^2\sinh^2\frac{\hbar\pi n}{mcL}\,.
\end{equation}
In the $c\rightarrow\infty$ limit we have
$\displaystyle\lim_{c\rightarrow\infty}E_n=\tfrac{\hbar^2\pi^2}{mL^2}2n^2$
and this is consistent with \eqref{RSCS}, \eqref{trigA_RSCS_para} and
\eqref{trigA_CS_Eorg}.

\section{Trigonometric $BC$ Types}

In this section we consider the CS and RSvD systems with the
trigonometric $BC$ type potentials.
The Jacobi polynomial and the Askey-Wilson polynomial play the role.

\subsection{Sutherland Systems}

The Hamiltonian is \eqref{H_CS} with the potential
\eqref{trigBC_CS_V}--\eqref{trigBC_CS_W}.

\subsubsection{Equilibrium positions of $n$-particle classical systems}

Let us consider a polynomial whose zeros give the equilibrium positions,
$f(\xi)=\prod_{j=1}^n\Bigl(\xi-\cos\bigl(2\tfrac{\pi}{L}\bar{q}_j\bigr)\Bigr)$.
Then \eqref{CS_equiveqW} can be converted to a differential equation
for $f(y)$, which determines the Jacobi polynomial. The result is
\begin{equation}
  \prod_{j=1}^n\Bigl(\xi-\cos\bigl(2\tfrac{\pi}{L}\bar{q}_j\bigr)\Bigr)
  =P_n^{(\alpha,\beta)\,{\rm monic}}(\xi)\,,\quad
  \alpha=\frac{\bar{g}_S+\bar{g}_L}{\bar{g}_M}-1\,,\
  \beta=\frac{\bar{g}_L}{\bar{g}_M}-1\,,
\end{equation}
where $P_n^{(\alpha,\beta)}(\xi)=2^{-n}\binom{\alpha+\beta+2n}{n}
P_n^{(\alpha,\beta)\,{\rm monic}}(\xi)$
is the Jacobi polynomial \cite{KS96}.

\subsubsection{Eigenfunctions of single-particle quantum mechanics}
\label{1particle_trigBC_CS}

Let us consider the single-particle case ($n=1$) and write $x=q_1$ and
$g=g_S+g_L$, $g'=g_L$.
The Hamiltonian \eqref{H_CS} has the P\"{o}schl-Teller potential \cite{PT33}
with a constant energy shift
\begin{equation}
  H=-\frac{\hbar^2}{2m}\frac{d^2}{dx^2}
  +\frac{\hbar^2\pi^2}{2mL^2}\Bigl(
  \frac{g(g-1)}{\sin^2\frac{\pi}{L}x}+\frac{g'(g'-1)}{\cos^2\frac{\pi}{L}x}
  \Bigr)
  -\frac{\hbar^2\pi^2}{2mL^2}(g+g')^2\,.
\end{equation}
By introducing a dimensionless variable $y$ \eqref{trig_CS_y},
$H$ can be written as
\begin{equation}
  H=\frac{\hbar^2\pi^2}{mL^2}\,\mathcal{H}\,,\quad
  \mathcal{H}=\frac12\Bigl(-\frac{d^2}{dy^2}+\frac{g(g-1)}{\sin^2y}
  +\frac{g'(g'-1)}{\cos^2y}-(g+g')^2\Bigr)\,.
\end{equation}
This $\mathcal{H}$ has parameters $g$ and $g'$ and we will denote
$\bm{\lambda}=(g,g')$.
In the following we will consider arbitrary positive parameters $g$ and $g'$.
Instead of $H\phi_n=E_n\phi_n$, let us consider
$\mathcal{H}\phi_n(y)=\mathcal{E}_n\phi_n(y)$ ($n=0,1,2,\ldots$),
where energies are related as $E_n=\frac{\hbar^2\pi^2}{mL^2}\mathcal{E}_n$.

Like in \S\ref{1particle_ratA_CS}, $\mathcal{H}$ is factorizable
\eqref{rat_CS_calH=AA}--\eqref{rat_CS_Ad}, where the prepotential with
parameters $\bm{\lambda}=(g,g')$ is
\begin{equation}
  \mathcal{W}(y\,;\bm{\lambda})=g\log\sin y+g'\log\cos y.
  \label{trigBC_CS_cW}
\end{equation}
The ground state wavefunction of $\mathcal{H}$ is \eqref{rat_CS_phi0},
\begin{equation}
  \phi_0(y\,;\bm{\lambda})\propto (\sin y)^g(\cos y)^{g'},\quad
  \mathcal{E}_0(\bm{\lambda})=0.
  \label{trig_BC_phi0}
\end{equation}
The Hamiltonian $\mathcal{H}$ is shape invariant \eqref{rat_CS_shapeinv} with
\begin{equation}
  \bm{\delta}=(1,1),\quad
  \mathcal{E}_1(\bm{\lambda})=2(g+g'+1).
  \label{trigBC_CS_E1}
\end{equation}
Let us define
$\mathcal{A}_s$, $\mathcal{H}_s$ and $\phi_{s,n}$ ($n\geq s\geq 0$)
by \eqref{rat_CS_A_rec}--\eqref{rat_CS_phi_rec}.
Then for $n\geq s\geq 0$ we obtain the consequence of the
shape-invariance \eqref{rat_CS_As}--\eqref{rat_CS_Adphi}
and \eqref{rat_CS_phi=Aphi}--\eqref{rat_CS_phi=phi}.
{}From \eqref{rat_CS_E_rec} and \eqref{trigBC_CS_E1} we obtain
\begin{equation}
  \mathcal{E}_n(\bm{\lambda})=2n(n+g+g').
  \label{trigBC_CS_E}
\end{equation}

The Rodrigues-type formula \eqref{rat_CS_phi=Adphi}--\eqref{rat_CS_phi=phi}
gives $\phi_n(y)\propto P_n^{(g-\frac12,g'-\frac12)}(\cos 2y)\phi_0(y)$.
This can be also understood as \eqref{rat_CS_tcalH}--\eqref{cHP=cEP},
\begin{align}
  \tilde{\mathcal{H}}&=-\frac12\frac{d^2}{dy^2}
  -\bigl(g\cot y-g'\tan y\bigr)\frac{d}{dy}\n
  &=-2\Bigl((1-\xi^2)\frac{d^2}{d\xi^2}
  -\bigl(g-g'+(g+g'+1)\xi\bigr)\frac{d}{d\xi}\Bigr)\,,\quad
  (\xi={\rm Re}z=\cos 2y).
\end{align}
Since the Jacobi polynomial satisfies
\begin{equation}
  \Bigl((1-\xi^2)\frac{d^2}{d\xi^2}
  -\bigl(\alpha-\beta+(\alpha+\beta+2)\xi\bigr)\frac{d}{d\xi}
  +n(n+\alpha+\beta+1)\Bigr)
  P_n^{(\alpha,\beta)}(\xi)=0,
\end{equation}
we obtain
\begin{equation}
  P_n(z\,;\bm{\lambda})\propto P_n^{(g-\frac12,g'-\frac12)}(\xi),\quad
  \mathcal{E}_n(\bm{\lambda})=2n(n+g+g').
\end{equation}

The energy spectrum of $H$ is
\begin{equation}
  E_n(\bm{\lambda})=\frac{\hbar^2\pi^2}{mL^2}2n(n+g+g')
  =\frac{\hbar^2\pi^2}{2mL^2}\Bigl((2n+g+g')^2-(g+g')^2\Bigr).
  \label{trigBC_CS_Eorg}
\end{equation}

\noindent
\underline{Remark}:
Similarly to Remark in \S\ref{1particle_ratBC_CS}, we have three other
prepotentials $\check{\mathcal{W}}^{[1]}$,
$\check{\mathcal{W}}^{[2]}$, $\check{\mathcal{W}}^{[3]}$,
\begin{align}
  \check{\mathcal{W}}^{[1]}\bigl(y\,;(g,g')\bigr)
  &=\mathcal{W}\bigl(y\,;(1-g,g')\bigr),\n
  \check{\mathcal{W}}^{[2]}\bigl(y\,;(g,g')\bigr)
  &=\mathcal{W}\bigl(y\,;(g,1-g')\bigr),\\
  \check{\mathcal{W}}^{[3]}\bigl(y\,;(g,g')\bigr)
  &=\mathcal{W}\bigl(y\,;(1-g,1-g')\bigr),\nonumber
\end{align}
together with the corresponding Hamiltonians $\check{\mathcal{H}}$:
\begin{align}
  \check{\mathcal{H}}^{[1]}(y\,;\bm{\lambda})
  &=\check{\mathcal{A}}^{[1]}(y\,;\bm{\lambda})^{\dagger}
  \check{\mathcal{A}}^{[1]}(y\,;\bm{\lambda})
  =\mathcal{H}(y\,;\bm{\lambda})+2(g-\tfrac12)(g'+\tfrac12),\n
  \check{\mathcal{H}}^{[2]}(y\,;\bm{\lambda})
  &=\check{\mathcal{A}}^{[2]}(y\,;\bm{\lambda})^{\dagger}
  \check{\mathcal{A}}^{[2]}(y\,;\bm{\lambda})
  =\mathcal{H}(y\,;\bm{\lambda})+2(g+\tfrac12)(g'-\tfrac12),\\
  \check{\mathcal{H}}^{[3]}(y\,;\bm{\lambda})
  &=\check{\mathcal{A}}^{[3]}(y\,;\bm{\lambda})^{\dagger}
  \check{\mathcal{A}}^{[3]}(y\,;\bm{\lambda})
  =\mathcal{H}(y\,;\bm{\lambda})+2(g+g'-1).\nonumber
\end{align}
The `ground' state of $\check{\mathcal{H}}^{[a]}$ ($a=1,2,3$) is
$\check{\phi}^{[a]}_0(y\,;\bm{\lambda})\propto
e^{\check{\mathcal{W}}^{[a]}(y\,;\bm{\lambda})}$ and
they are square integrable for $g<\frac32$ and/or $g'<\frac32$.
Note that $\phi_0(y\,;\bm{\lambda})$ \eqref{trig_BC_phi0} is square
integrable for $g,g'>-\frac12$. 
These four `ground' states define four sectors of this system.
Usually we consider only one of them. 
Note that $\mathcal{A}(y\,;g)\check{\phi}^{[a]}_0(y\,;g)$ is square
integrable for $g\leq\frac12$ and/or $g'\leq\frac12$.
{}From the Rodrigues-type formula and the recurrence relation of the
energy, we obtain
\begin{align}
  \check{\phi}^{[1]}_n(y\,;\bm{\lambda})\propto
  P_n^{(\frac12-g,g'-\frac12)}(\cos 2y)
  \check{\phi}^{[1]}_0(y\,;\bm{\lambda}),
  \quad \check{\mathcal{E}}^{[1]}_n(\bm{\lambda})=2n(n+1-g+g'),\n
  \check{\phi}^{[2]}_n(y\,;\bm{\lambda})\propto
  P_n^{(g-\frac12,\frac12-g')}(\cos 2y)
  \check{\phi}^{[2]}_0(y\,;\bm{\lambda}),
  \quad \check{\mathcal{E}}^{[2]}_n(\bm{\lambda})=2n(n+1+g-g'),
  \label{trigBC_CS_checkphi}\\
  \check{\phi}^{[3]}_n(y\,;\bm{\lambda})\propto
  P_n^{(\frac12-g,\frac12-g')}(\cos 2y)
  \check{\phi}^{[3]}_0(y\,;\bm{\lambda}),
  \quad \check{\mathcal{E}}^{[3]}_n(\bm{\lambda})=2n(n+2-g-g').\nonumber
\end{align}
The corresponding energy spectra of $H$ are
\begin{align}
  \check{E}^{[1]}_n(\bm{\lambda})&=\frac{\hbar^2\pi^2}{2mL^2}\Bigl(
  (2n+1-g+g')^2-(g+g')^2\Bigr),\n
  \check{E}^{[2]}_n(\bm{\lambda})&=\frac{\hbar^2\pi^2}{2mL^2}\Bigl(
  (2n+1+g-g')^2-(g+g')^2\Bigr),
  \label{trigBC_CS_E'org}\\
  \check{E}^{[3]}_n(\bm{\lambda})&=\frac{\hbar^2\pi^2}{2mL^2}\Bigl(
  (2n+2-g-g')^2-(g+g')^2\Bigr).
  \nonumber
\end{align}
Therefore, for $g<\frac32$ and/or $g'<\frac32$, we have these sectors.
$\phi_0(y\,;\bm{\lambda})$ is the lowest energy state of $H$ for 
$g,g'\geq\frac32$ or $-\frac12<g,g'\leq\frac12$.
In the $g,g'\rightarrow 0$ (or 1) limit, all of these sectors contribute
and these eigenfunctions reduce to those in \S\ref{1particle_trigA_CS} 
due to the identities,
\begin{align}
  \phi_n\bigl(y\,;(g\rightarrow 0,g'\rightarrow 0)\bigr)&\propto
  P_n^{(-\frac12,-\frac12)}(\cos 2y)\propto \cos 2ny,\\
  \check{\phi}^{[1]}_n\bigl(y\,;(g\rightarrow 0,g'\rightarrow 0)\bigr)&\propto
  P_n^{(\frac12,-\frac12)}(\cos 2y)\sin y\propto \sin (2n+1)y,
  \label{phihat1->0}\\
  \check{\phi}^{[2]}_n\bigl(y\,;(g\rightarrow 0,g'\rightarrow 0)\bigr)&\propto
  P_n^{(-\frac12,\frac12)}(\cos 2y)\cos y\propto \cos (2n+1)y,
  \label{phihat2->0}\\
  \check{\phi}^{[3]}_n\bigl(y\,;(g\rightarrow 0,g'\rightarrow 0)\bigr)&\propto
  P_n^{(\frac12,\frac12)}(\cos 2y)\sin y\cos y\propto \sin 2(n+1)y.
\end{align}
(The two sets \eqref{phihat1->0} and \eqref{phihat2->0} are excluded by the
periodic boundary condition in $x$.)

\subsection{Ruijsenaars-Schneider-van Diejen Systems}

The Hamiltonian is \eqref{H_RS} with the potential
\eqref{trigBC_RS_v}--\eqref{trigBC_RS_w}.

\subsubsection{Equilibrium positions of $n$-particle classical systems}

Let us consider a polynomial whose zeros give the equilibrium positions,
$f(\xi)=\prod_{j=1}^n\Bigl(\xi-\cos\bigl(2\tfrac{\pi}{L}\bar{q}_j\bigr)\Bigr)$.
Then \eqref{equiveqBA} can be converted to a functional equation
for $f(y)$.
We can show that the solutions of this functional equation satisfy the 
three-term recurrence which agrees with that of the Wilson polynomials. 
The result is \cite{OS2,vD04,OS5}(see also \cite{ILR04})
\begin{equation}
  \prod_{j=1}^n\Bigl(\xi-\cos\bigl(2\tfrac{\pi}{L}\bar{q}_j\bigr)\Bigr)
  =p_n^{\rm monic}\bigl(\xi\,;
  e^{-\frac{2\pi\bar{g}_1}{mcL}},e^{-\frac{2\pi\bar{g}_2}{mcL}},
  -e^{-\frac{2\pi\bar{g}'_1}{mcL}},-e^{-\frac{2\pi\bar{g}'_2}{mcL}}
  \bigm|e^{-\frac{2\pi\bar{g}_0}{mcL}}\bigr),
\end{equation}
where $p_n(\xi\,;a_1,a_2,a_3,a_4|q)=2^n(a_1a_2a_3a_4q^{n-1};q)_n\,
p_n^{\rm monic}(\xi\,;a_1,a_2,a_3,a_4|q)$ is the Askey - Wilson polynomial
\cite{KS96}.
Note that $e^{-\frac{2\pi\bar{g}_0}{mcL}}$ etc. are formally expressed as
$e^{-\frac{2\pi\hbar g_0}{mcL}}=q^{g_0}$ etc. by using $q$ in \eqref{q}.

\subsubsection{Eigenfunctions of single-particle quantum mechanics}
\label{1particle_trigBC_RS}

Let us consider the single-particle case ($n=1$).
The potential $V_1(q)$ is $V_1(q)=w(q_1)$. Let us write $x=q_1$.
The Hamiltonian \eqref{H_RS} becomes \eqref{rat_RS_H} with $w(x)$ in
\eqref{trigBC_RS_w}.
By using $y$ \eqref{trig_CS_y}, discussion goes parallel to that in
\S\ref{1particle_ratBC_RS}, but the variable $z$ \eqref{trig_CS_z} is
more suitable. So we will reformulate by using $z$ and $q$ \eqref{q}.
By introducing a dimensionless variable $z$ \eqref{trig_CS_z} and
a rescaled potential $V(z)$,
\begin{equation}
  V\bigl(z\,;(a_1,a_2,a_3,a_4),q\bigr)
  =\frac{(1-a_1z)(1-a_2z)(1-a_3z)(1-a_4z)}{(1-z^2)(1-qz^2)}\,,
  \label{trigBC_RS_V}
\end{equation}
$w(x)$ and $H$ are expressed as
\begin{gather}
  w(x)^*=q^{-\frac12(g_1+g_2+g'_1+g'_2)}
  V(z\,;(q^{g_1},q^{g_2+\frac12},-q^{g'_1},-q^{g'_2+\frac12}),q),\\
  H=mc^2q^{-\frac12(g_1+g_2+g'_1+g'_2)}\mathcal{H}.
\end{gather}
Here $\mathcal{H}$ is defined by
\begin{equation}
  \mathcal{H}=\frac12\Bigl(\sqrt{V(z)}\,q^{D_z}\sqrt{V(z)^*}
  +\sqrt{V(z)^*}\,q^{-D_z}\sqrt{V(z)}-V(z)-V(z)^*\Bigr)\,,
\end{equation}
where $V(z)$ is $V(z;\bm{\lambda},q)$ \eqref{trigBC_RS_V} with
$\bm{\lambda}=(a_1,a_2,a_3,a_4)=
(q^{g_1},q^{g_2+\frac12},-q^{g'_1},-q^{g'_2+\frac12})$.
In the following we will consider the parameters in the range
$-1<a_1,a_2,a_3,a_4<1$ and $0<q<1$.
Instead of $H\phi_n=E_n\phi_n$, let us consider the rescaled equation
$\mathcal{H}\phi_n(z)=\mathcal{E}_n\phi_n(z)$ $(n=0,1,2,\ldots)$,
where energies are related as
$E_n=mc^2q^{-\frac12(g_1+g_2+g'_1+g'_2)}\mathcal{E}_n$.

Like in \S\ref{1particle_ratBC_RS}, $\mathcal{H}$ is factorizable:
\begin{align}
  \mathcal{H}&=\mathcal{H}(z\,;\bm{\lambda},q)
  =\mathcal{A}(z\,;\bm{\lambda},q)^{\dagger}\mathcal{A}(z\,;\bm{\lambda},q)\,,
  \label{trig_RS_calH=AA}\\
  \mathcal{A}&=\mathcal{A}(z\,;\bm{\lambda},q)\eqdef\frac{1}{\sqrt{2}}\Bigl(
  q^{\frac12D_z}\sqrt{V(z\,;\bm{\lambda},q)^*}
  -q^{-\frac12D_z}\sqrt{V(z\,;\bm{\lambda},q)}\,\Bigr)\,,\\
  \mathcal{A}^{\dagger}&=\mathcal{A}(z\,;\bm{\lambda},q)^{\dagger}
  \eqdef\frac{1}{\sqrt{2}}\Bigl(
  \sqrt{V(z\,;\bm{\lambda},q)}\,q^{\frac12D_z}
  -\sqrt{V(z\,;\bm{\lambda},q)^*}\,q^{-\frac12D_z}\Bigr)\,.
\end{align}
The ground state of $\mathcal{H}$ is annihilated by $\mathcal{A}$,
\begin{equation}
  \phi_0(z)\propto \biggl|
  \frac{(z^2;q)_{\infty}}{(a_1z,a_2z,a_3z,a_4z;q)_{\infty}}\biggr|\,.
\end{equation}
The Hamiltonian $\mathcal{H}$ is shape invariant but slightly different from
the previous form \eqref{rat_CS_shapeinv}
\begin{equation}
  \mathcal{A}(z\,;\bm{\lambda},q)\mathcal{A}(z\,;\bm{\lambda},q)^{\dagger}
  =q^{2\delta'}\mathcal{A}(z\,;q^{\delta}\bm{\lambda},q)^{\dagger}
  \mathcal{A}(z\,;q^{\delta}\bm{\lambda},q)+\mathcal{E}_1(\bm{\lambda},q)\,,
  \label{trig_RS_shapeinv}
\end{equation}
with\footnote{
If we include a factor $(a_1a_2a_3a_4)^{-\frac12}$ into $V(z)$
(namely $w(x)^*$), then $\delta'$ becomes $0$.
}
\begin{equation}
  \delta=\frac12,\quad\delta'=-\frac12,\quad
  \mathcal{E}_1(\bm{\lambda},q)=\tfrac12(q^{-1}-1)(1-a_1a_2a_3a_4)\,.
  \label{trigBC_RS_E1}
\end{equation}
This $\delta$ is consistent with $\bm{\delta}$ in \eqref{trigBC_CS_E1}
because of \eqref{RSCS} and \eqref{trigBC_RSCS_para}.

Starting from $\mathcal{A}_0=\mathcal{A}$, $\mathcal{H}_0=\mathcal{H}$
and $\phi_{0,n}=\phi_n$, let us define $\mathcal{A}_s$, $\mathcal{H}_s$
and $\phi_{s,n}$ ($n\geq s\geq 0$) recursively:
\begin{align}
  \mathcal{A}_{s+1}(z\,;\bm{\lambda},q)&\eqdef
  q^{\delta'}\mathcal{A}_s(z\,;q^{\delta}\bm{\lambda},q)\,,
  \label{trig_RS_A_rec}\\
  \mathcal{H}_{s+1}(z\,;\bm{\lambda},q)&\eqdef
  \mathcal{A}_s(z\,;\bm{\lambda},q)\mathcal{A}_s(z\,;\bm{\lambda},q)^{\dagger}
  +\mathcal{E}_s(\bm{\lambda},q)\,,\\
  \phi_{s+1,n}(z\,;\bm{\lambda},q)&\eqdef
  \mathcal{A}_s(z\,;\bm{\lambda},q)\phi_{s,n}(z\,;\bm{\lambda},q)\,.
  \label{trig_RS_phi_rec}
\end{align}
As a consequence of the shape invariance \eqref{trig_RS_shapeinv},
we obtain for $n\geq s\geq 0$,
\begin{align}
  &\mathcal{A}_s(z\,;\bm{\lambda},q)=q^{s\delta'}
  \mathcal{A}(z\,;q^{s\delta}\bm{\lambda},q)\,,\\
  &\mathcal{H}_s(z\,;\bm{\lambda},q)
  =\mathcal{A}_s(z\,;\bm{\lambda},q)^{\dagger}\mathcal{A}_s(z\,;\bm{\lambda},q)
  +\mathcal{E}_s(\bm{\lambda},q)
  =q^{2s\delta'}\mathcal{H}(z\,;q^{s\delta}\bm{\lambda},q)
  +\mathcal{E}_s(\bm{\lambda},q)\,,
  \label{trig_RS_H_rec}\\
  &\mathcal{E}_{s+1}(\bm{\lambda},q)
  =\mathcal{E}_s(\bm{\lambda},q)
  +q^{2s\delta'}\mathcal{E}_1(q^{s\delta}\bm{\lambda},q)\,,
  \label{trig_RS_E_rec}\\
  &\mathcal{H}_s(z\,;\bm{\lambda},q)\phi_{s,n}(z\,;\bm{\lambda},q)
  =\mathcal{E}_n(\bm{\lambda},q)\phi_{s,n}(z\,;\bm{\lambda},q)\,,\\
  &\mathcal{A}_s(z\,;\bm{\lambda},q)\phi_{s,s}(z\,;\bm{\lambda},q)=0\,,\\
  &\mathcal{A}_s(z\,;\bm{\lambda},q)^{\dagger}\phi_{s+1,n}(z\,;\bm{\lambda},q)
  =\bigl(\mathcal{E}_n(\bm{\lambda},q)-\mathcal{E}_s(\bm{\lambda},q)\bigr)
  \phi_{s,n}(z\,;\bm{\lambda},q)\,.
  \label{trig_RS_Adphi}
\end{align}
{}From \eqref{trig_RS_phi_rec} and \eqref{trig_RS_Adphi} we obtain formulas,
\begin{align}
  \phi_{s,n}(z\,;\bm{\lambda},q)&=\mathcal{A}_{s-1}(z\,;\bm{\lambda},q)\cdots
  \mathcal{A}_1(z\,;\bm{\lambda},q)
  \mathcal{A}_0(z\,;\bm{\lambda},q)\phi_n(z\,;\bm{\lambda},q)\,,\\
  \phi_n(z\,;\bm{\lambda},q)&=
  \frac{\mathcal{A}_0(z\,;\bm{\lambda},q)^{\dagger}}
       {\mathcal{E}_n(\bm{\lambda},q)-\mathcal{E}_0(\bm{\lambda},q)}\,
  \frac{\mathcal{A}_1(z\,;\bm{\lambda},q)^{\dagger}}
       {\mathcal{E}_n(\bm{\lambda},q)-\mathcal{E}_1(\bm{\lambda},q)}\cdots
  \frac{\mathcal{A}_{n-1}(z\,;\bm{\lambda},q)^{\dagger}}
       {\mathcal{E}_n(\bm{\lambda},q)-\mathcal{E}_{n-1}(\bm{\lambda},q)}\,
  \phi_{n,n}(z\,;\bm{\lambda},q)\,,
  \label{trig_RS_phi=Adphi}
\end{align}
and from \eqref{trig_RS_H_rec} we have
\begin{equation}
  \phi_{n,n}(z\,;\bm{\lambda},q)\propto
  \phi_0(z\,;q^{n\delta}\bm{\lambda},q).
\end{equation}
{}From \eqref{trig_RS_E_rec} and \eqref{trigBC_RS_E1}, we obtain
\begin{equation}
  \mathcal{E}_n(\bm{\lambda},q)=\tfrac12(q^{-n}-1)(1-a_1a_2a_3a_4q^{n-1}).
  \label{trigBC_RS_E}
\end{equation}

By similarity transformation in terms of the ground state wavefunction,
let us define $\tilde{\mathcal{H}}$,
\begin{align}
  \tilde{\mathcal{H}}&=\phi_0(z\,;\bm{\lambda},q)^{-1}\circ
  \mathcal{H}\circ\phi_0(z\,;\bm{\lambda},q)
  =\frac12\Bigl(V(z)q^{D_z}+V(z)^*q^{-D_z}
  -V(z)-V(z)^*\Bigr),
  \label{trig_RS_tcalH}\\
  &=BC,\quad
  B=V(z)q^{\frac12D_z}-V(z)^*q^{-\frac12D_z},\quad
  C=\tfrac12\bigl(q^{\frac12D_z}-q^{-\frac12D_z}\bigr),
  \label{tirg_RS_BC}
\end{align}
and consider $\phi_n(z\,;\bm{\lambda},q)=P_n(z\,;\bm{\lambda},q)
\phi_0(z\,;\bm{\lambda},q)$, where $P_n(z\,;\bm{\lambda},q)$ satisfies
\begin{equation}
  \tilde{\mathcal{H}}(z\,;\bm{\lambda},q)P_n(z\,;\bm{\lambda},q)
  =\mathcal{E}_n(\bm{\lambda},q)P_n(z\,;\bm{\lambda},q).
\end{equation}
This means that $P_n(z\,;\bm{\lambda},q)$ is the Askey-Wilson polynomial
\begin{equation}
  P_n(z;\bm{\lambda},q)\propto p_n({\rm Re}z\,;a_1,a_2,a_3,a_4|q),\quad
  \mathcal{E}_n(\bm{\lambda},q)=\tfrac12(q^{-n}-1)(1-a_1a_2a_3a_4q^{n-1}).
\end{equation}

The energy spectrum of $H$ is
\begin{align}
  E_n&=mc^2q^{-\frac12(g_1+g_2+g'_1+g'_2)}
  \tfrac12(q^{-n}-1)(1-q^{n+g_1+g_2+g'_1+g'_2})\n
  &=2mc^2\sinh\frac{\hbar\pi n}{mcL}
  \sinh\frac{\hbar\pi(n+g_1+g_2+g'_1+g'_2)}{mcL}.
\end{align}
In the $c\rightarrow\infty$ limit we have
$\displaystyle\lim_{c\rightarrow\infty}E_n
=\frac{\hbar^2\pi^2}{mL^2}2n(n+g_1+g_2+g'_1+g'_2)$
and this is consistent with \eqref{RSCS}, \eqref{trigBC_RSCS_para} and
\eqref{trigBC_CS_Eorg}.

\section{Summary and Comments}

We have reviewed some interesting properties of the
Calogero-Sutherland-Moser systems and the Ruijsenaars-Schneider-van Diejen 
systems with the rational and trigonometric potentials.
The equilibrium positions of classical multi-particle systems and the 
eigenfunctions of single-particle quantum  mechanics are described by
the same orthogonal polynomials: the Hermite, Laguerre, Jacobi, 
continuous Hahn, Wilson and Askey-Wilson polynomials.
This interesting property was obtained as a result of explicit
computation and we do not know any deeper reason or meaning behind it.
The CSM and RSvD systems admit elliptic potentials and finding
eigenfunctions of such elliptic systems is a good challenge.
If this property is inherited by the elliptic cases, study of classical 
equilibrium positions may shed light on the quantum problem of finding 
eigenfunctions, which is quite non-trivial.

When we discuss the Hamiltonians of these single-particle quantum
mechanics, we have emphasized factorization, shape invariance and 
construction of the isospectral Hamiltonians. Although the examples
given in this note are rational and trigonometric potentials, this
method and idea could be applied to a wider class of potentials, e.g. 
elliptic potential.
In ordinary quantum mechanics there is the Crum's theorem \cite{Crum55},
which states a construction of the associated isospectral Hamiltonians
$\mathcal{H}_s$ and their eigenfunctions $\phi_{s,n}$ for general systems
without invariance. The construction of $\mathcal{H}_s$ and
$\phi_{s,n}$ given in this note for `discrete' cases needs shape
invariance. A `discrete' analogue of the Crum's theorem, namely similar
construction without shape invariance, would be very helpful, if it exists.

It should be mentioned that in the discussion of various
`eigenfunctions', the function theory aspects are more emphasized than
the ordinary quantum mechanical considerations in
\S\ref{1particle_ratBC_CS}, \S\ref{1particle_trigA_CS} and
\S\ref{1particle_trigBC_CS}.

\section*{Acknowledgements}

S.O. would like to thank M.~Noumi and K.~Takasaki,
the organizers of the workshop, for giving him an opportunity to talk
and for a financial support.
S.O. and R.S. are supported in part by Grant-in-Aid for Scientific
Research from the Ministry of Education, Culture, Sports, Science and
Technology, No.13135205 and No. 14540259, respectively.


\end{document}